\documentclass[%aip,
 amsmath,amssymb,
preprint,%
nofootinbib
]{revtex4}

\usepackage{bm}
\usepackage{makecell}
\usepackage{multirow}
\usepackage{graphicx}
\usepackage{epstopdf}
\usepackage{tablefootnote}
\usepackage{subfigure}
\usepackage{amsmath}
\usepackage{caption}
\usepackage{color}

\usepackage{lscape}

\newcommand{\B}{{\cal B}}
\newcommand{\A}{{\cal A}}
\newcommand{\ov}{\overline }
\newcommand{\Bc}{(\mathcal{B}_c)}
\newcommand{\Boct}{(\mathcal{B}_8)}

\newcommand{\CP}{{\it CP}~}

\begin{document}

\title{Hadronic Weak Decays of Charmed Baryons\\ in the Topological Diagrammatic Approach: An Update}

\author{Hai-Yang Cheng}
\affiliation{Institute of Physics, Academia Sinica, Taipei, Taiwan 11529, Republic of China}

\author{Fanrong Xu}
\email{fanrongxu@jnu.edu.cn}
\author{Huiling Zhong}
 \affiliation{Department of Physics, College of Physics $\&$ Optoelectronic Engineering, Jinan University, Guangzhou 510632, P.R. China}

%\affiliation{ 
%Authors' institution and/or address%\\This line break forced with \textbackslash\textbackslash
%}%

\small

\vskip 0.5cm
\begin{abstract}
\small
\vskip 0.5cm
There exist two distinct ways in realizing the approximate SU(3) flavor symmetry of QCD to describe the two-body nonleptonic decays of charmed baryons: the irreducible SU(3) approach (IRA) and the topological diagram approach (TDA). The TDA has the advantage that it is more intuitive, graphic and easier to implement model calculations. We perform a global fit to the currently available data of two-body charmed baryon decays within the framework of the TDA and IRA. The number of the minimum set of tensor invariants in the IRA and the topological amplitudes in the TDA is the same, namely, five in the tree-induced amplitudes and four in the penguin amplitudes. 
Since we employ the new LHCb measurements to fix the sign ambiguity of the decay parameters $\beta$ and $\gamma$, the fit results for  the magnitudes of $S$- and $P$-wave amplitudes and their phase shift $\delta_P-\delta_S$ in both the TDA and IRA are more
trustworthy than our previous analyses with uncertainties substantially improved.  These results  can be tested in the near future. The perspective of having direct \CP violation in the charmed baryon sector at the per mille level is briefly discussed.

\end{abstract}

%Use showkeys class option if keyword
\maketitle

\section{Introduction}
\label{intro}
In the past few years, the experimental and theoretical progresses in the study of hadronic decays of charmed baryons are very impressive. On the experimental side, more than 35 measurements of  branching fractions and decay parameters have been accumulated. On the theory aspect, there were many approaches developed in 1990s such as the relativistic quark model, the pole model and current algebra (for a review, see 
Ref. ~\cite{Cheng:2021qpd}).
Besides the dynamical model calculations, a very promising approach is to use the approximate SU(3) flavor symmetry of QCD to describe the two-body nonleptonic decays of charmed baryons.  There exist two distinct ways in realizing the flavor symmetry: the irreducible SU(3) approach (IRA) and the topological diagram approach (TDA). They provide a powerful tool for a model-independent analysis. Among them, the IRA has become very popular in recent years. In the IRA, SU(3) tensor invariants are constructed through the short-distance effective Hamiltonian, while in the TDA, the topological diagrams are classified according to the topologies in the flavor flow of weak decay diagrams with all strong-interaction effects included implicitly.  

Within the framework of the IRA, two-body nonleptonic decays of charmed baryons were first analyzed in Refs. \cite{Savage,Verma}.  After 2014, this approach became rather popular. 
However, the early studies of the IRA had overlooked the fact that charmed baryon decays are governed by several different partial-wave amplitudes which have distinct kinematic and dynamic effects. That is, $S$- and $P$-waves were not distinguished in the early analyses and the IRA amplitudes are fitted only to the measured rates.
After the pioneer work in Ref. \cite{Geng:2019xbo}, it became a common practice to perform a global fit of both $S$- and $P$-wave parameters to the data of branching fractions and decay asymmetries \cite{Geng:2019awr,Geng:2020zgr,Huang:2021aqu,Zhong:2022exp,Xing:2023dni}.
Just like the case of hyperon decays, non-trivial relative strong phases between $S$- and $P$-wave amplitudes may exist, but they were usually not considered in realistic model calculations of the decay asymmetry $\alpha$. 

The first analysis of two-body nonleptonic decays of antitriplet charmed baryons 
$\B_c(\bar 3)\to \B(8) P(8+1)$ within the framework of the TDA was performed by Kohara \cite{Kohara:1991ug}. A subsequent study was given by Chau, Cheng and Tseng (CCT) in Ref. \cite{Chau:1995gk} followed by some recent analyses in the TDA \cite{He:2018joe,Hsiao:2021nsc,Zhao:2018mov,Hsiao:2020iwc}. Unlike the IRA, global fits to the rates and decay asymmetries in the TDA were not available until recently.  

Although the TDA has been applied very successfully to charmed meson decays \cite{CC,Cheng:2016,Cheng:2024hdo}, its application to charmed baryon decays is more complicated than the IRA. As stressed in Ref. \cite{He:2018joe}, it is easy to determine the independent amplitudes in the IRA, while the TDA gives some redundancy. Some of the amplitudes are not independent and therefore should be absorbed into other amplitudes. Nevertheless, the TDA has the advantage that it is more intuitive, graphic and easier to implement model calculations.  The extracted topological amplitudes by fitting to the available data will enable us to probe the relative importance of different underlying decay mechanisms, and to relate one process to another at the topological amplitude level. 

In this work we will focus on the hadronic weak decays of antitriplet charmed baryons into a octet baryon and a pseudoscalar meson: ${\cal B}_c(\bar 3)\to {\cal B}(8)P(8+1)$. Its general decay amplitude reads
\begin{eqnarray}
\label{eq:A&B}
M(\B_c\to \B_f+P)=i\bar u_f(A-B\gamma_5)u_c,
\end{eqnarray}
where $A$ and $B$ correspond to the parity-violating $S$-wave and parity-conserving $P$-wave amplitudes, respectively. Three Lee-Yang decay parameters $\alpha$, $\beta$ and $\gamma$ are of particular interest. They are defined by 
\begin{equation}
\begin{split}
& \alpha=\frac{2\kappa |A^*B|\cos(\delta_P-\delta_S)}{|A|^2+\kappa^2 |B|^2},~~
\beta=\frac{2\kappa |A^*B|\sin(\delta_P-\delta_S)}{|A|^2+\kappa^2 |B|^2},~~
\gamma=\frac{|A|^2-\kappa^2 |B|^2}{|A|^2+\kappa^2 |B|^2},
\end{split}
\label{eq:decayparameter}
\end{equation}
with  $\kappa=p_c/(E_f+m_f)=\sqrt{(E_f-m_f)/(E_f+m_f)}$ and $\alpha^2+\beta^2+\gamma^2=1$. For completeness, the decay rate is given by
\begin{equation}
\label{eq:Gamma}
\Gamma = \frac{p_c}{8\pi}\frac{(m_i+m_f)^2-m_P^2}{m_i^2}\left(|A|^2
+ \kappa^2|B|^2\right). 
\end{equation}

To analyze these decays in the framework of the TDA or IRA, we need the inputs of the experimental measurements of branching fractions and decay parameters to fix the unknown coefficients in the TDA and IRA. However, all the current global fits to the data of $\Gamma$ and $\alpha$ encounter two major issues. First of all, there is a sign ambiguity for the decay parameter $\beta$ as $\alpha$ is proportional to $\cos(\delta_P-
\delta_S)$, while $\beta$ to $\sin(\delta_P-\delta_S)$. In other words, a measurement of $\alpha$ alone does not suffice to fix the phase shift between $S$- and $P$-wave amplitudes. Second, there is also a sign ambiguity for the decay parameter $\gamma$ as generally there exist two different set of solutions for $A$ and $B$ based on the input of $\Gamma$. One of the solutions yields $|A|>\kappa|B|$, whereas the other with  $|A|<\kappa|B|$, corresponding to positive and negative $\gamma$, respectively. It is obvious from Eqs. (\ref{eq:decayparameter}) and (\ref{eq:Gamma}) that one needs both information of $\Gamma$ and $\gamma$ to fix $|A|$ and $|B|$. 

%%%%%%%%%%%%%%%%%%%%%%%%%%%%%%%
\begin{table}[t]\footnotesize
\caption{LHCb measurements of the decay parameters for $\Lambda_c^+\to\Lambda \pi^+$  $\Lambda_c^+\to \Lambda K^+$ and $\Lambda_c^+\to pK_S^0$ decays \cite{LHCb:2024tnq}. }
 \label{tab:LHCb}
\begin{center} 
\begin{tabular}{l c c c  c}
\hline \hline
Decay  & $\alpha$ & $\beta$ & $\gamma$ & $\Delta$ (radian)\\
\hline 
$\Lambda_c^+\to\Lambda \pi^+$ & $-0.782\pm0.009\pm0.004$  & $0.368\pm0.019\pm0.008$    &  $0.502\pm0.016\pm0.006$ & $0.633\pm0.036\pm0.013$ \\
$\Lambda_c^+\to\Lambda K^+$ & $-0.569\pm0.059\pm0.028$  & $0.35\pm0.12\pm0.04$    &  $-0.743\pm0.067\pm0.024$ & $2.70\pm0.17\pm0.04$ \\
$\Lambda_c^+\to pK_S^0$ & $-0.744\pm0.012\pm0.009$ & -- & --  & --\\
\hline \hline
\end{tabular}
\end{center} 
\end{table}
%%%%%%%%%%%%%%%%%%%%%%%%%%%%%%%%%%%

Very recently, LHCb has precisely measured the $\beta$ and $\gamma$ parameters of $\Lambda_c^+\to \Lambda \pi^+$ and $\Lambda_c^+\to \Lambda K^+$  decays for the first time \cite{LHCb:2024tnq} (see Table \ref{tab:LHCb}). A quantity $\Delta={\rm arg}(H_+/H_-)$, which is the phase difference between the  two helicity amplitudes, was measured by LHCb with the results shown in Table \ref{tab:LHCb}. The decay parameters are then determined by
\begin{equation}
\beta=\sqrt{1-\alpha^2}\,\sin\Delta, \qquad \gamma=\sqrt{1-\alpha^2}\,\cos\Delta.
\end{equation}
These measurements will enable us to fix the phase shift $\delta_P-\delta_S$ and pick up the right solution for $S$- and $P$-wave amplitudes.  By the end of 2023, there were 30 experimental data of $\B$ and $\alpha$. Right now we have totally 38 data available, including the new measurements of $\beta$ and $\gamma$ decay parameters by LHCb and new results of the branching fractions of $\Xi_c^0\to\Xi^0 \pi^0, 
~\Xi^0 \eta, ~\Xi^0 \eta'$ and the decay asymmetry $\alpha(\Xi_c^0\to\Xi^0\pi^0)$ by Belle-II \cite{Belle-II:2024jql} (see Table \ref{tab:expandave}).  These data allow to perform more sensible global fits.

One of the main purposes of this work is to pave the way towards the exploration of  {\it CP}-violating effects in the charmed baryon sector.
Therefore, we introduce various penguin contractions and write down the topological
penguin amplitudes in both TDA and IRA which transform as ${\bf 3}$ and are proportional to the CKM matrix element $\lambda_b=V_{cb}^*V_{ub}$. These amplitudes are denoted by  ${\cal A}^{\lambda_b}_{\rm TDA}$ and ${\cal A}^{\lambda_b}_{\rm IRA}$, respectively, below. In the TDA, 
the coefficients are related to penguin and tree topologies and hence they have clear graphic picture. The coefficients in ${\cal A}^{\lambda_b}_{\rm TDA}$ or ${\cal A}^{\lambda_b}_{\rm IRA}$ are unknown due to the lack of experiments on \CP asymmetries. Nevertheless, they can be related to the 
tree topologies in ${\cal A}^{\rm tree}_{\rm TDA}$ or ${\cal A}^{\rm tree}_{\rm IRA}$
through final-state rescattering effects, a task which we will present in the future. Once  the coefficients are determined, we are allowed to investigate the rich {\it CP}-violating phenomenology in charmed baryon decays.  Moreover, \CP asymmetries are allowed to enhance 
from  $10^{-4}$ to the per mille level. This is the main idea behind our work on direct \CP asymmetries. 

In this work we will also update our previous analyses \cite{Zhong:2024qqs,Zhong:2024zme} by incorporating the new data for global fits. In particular, we pay special attention to the newly measured decay parameters $\beta$ and $\gamma$. Compared to the  previous ones \cite{Zhong:2024qqs,Zhong:2024zme}, the analysis in this work has two major improvements and innovations: (i) Penguin diagrams are introduced and topological penguin amplitudes are constructed, setting the stage  for further study of \CP violation.
(ii)
The sign ambiguities of the $\beta$ and $\gamma$ parameters in charmed baryon decays are resolved thanks to the new LHCb measurements of these Lee-Yang parameters in $\Lambda_c^+\to \Lambda \pi^+$ and $\Lambda_c^+\to \Lambda K^+$. Moreover, our expressions for the the phase shift $\delta_P-\delta_S$
in terms of $\alpha$ and $\beta$
and the phase difference $\Delta$ between the two helicity amplitudes in terms of $\beta$ and $\gamma$
have the great advantage that they are free of any possible ambiguity of $\pm \pi$ (see Eqs. (\ref{eq:phase}) and (\ref{eq:Delta}) below). 

The layout of this work is as follows. In Sec. II we discuss  the general expression of the decay amplitudes in the TDA, we show that the number of independent amplitudes can be reduced through the K\"orner-Pati-Woo theorem and the removal of redundancy. The equivalence of the TDA and IRA is also explicitly demonstrated. Sec. III is devoted to the numerical analysis and fitting results. The perspective of direct \CP violation in the charmed baryon sector is briefly discussed in Sec. IV.
Sec. V comes to our conclusions. Currently available experimental data are collected in the Appendix. 

\section{Formulism}
\label{sec-1}
Since baryons are made of three quarks in contrast to two quarks for the mesons, the application of the TDA to the baryon case will inevitably lead to some complications. As shown explicitly in Ref. \cite{Kohara:1997nu}, physics is independent of the convention one chooses
for the wave functions of the octet baryons.
We prefer to use the bases $\psi^k(8)_{A_{12}}$ and $\psi^k(8)_{S_{12}}$ for octet baryons as they are orthogonal to each other:
\begin{eqnarray}
|\psi^k(8)_{A_{12}}\rangle &=& \sum_{q_a,q_b,q_c}|[q_aq_b]q_c\rangle \langle[q_aq_b]q_c
|\psi^k(8)_{A_{12}}\rangle, \nonumber \\
|\psi^k(8)_{S_{12}}\rangle &=& \sum_{q_a,q_b,q_c}|\{q_aq_b\}q_c\rangle \langle\{q_aq_b\}q_c
|\psi^k(8)_{S_{12}}\rangle, 
\end{eqnarray}
denoting the octet baryon states that are antisymmetric and symmetric in the first two quarks, respectively. Hence,
\begin{eqnarray} \label{eq:wf8}
|{\cal B}^{m,k}(8)\rangle=a \,|\chi^m(1/2)_{A_{12}}\rangle|\psi^k(8)_{A_{12}}\rangle+ b\, |\chi^m(1/2)_{S_{12}}\rangle|\psi^k(8)_{S_{12}}\rangle
\end{eqnarray}
with $|a|^2+|b|^2=1$,  where $\chi^m(1/2)_{A,S}$ are the spin parts of the wave function defined in Eq. (23) of Ref. \cite{Chau:1995gk}.

\begin{figure}[t]
\centering
\includegraphics[scale=0.66]{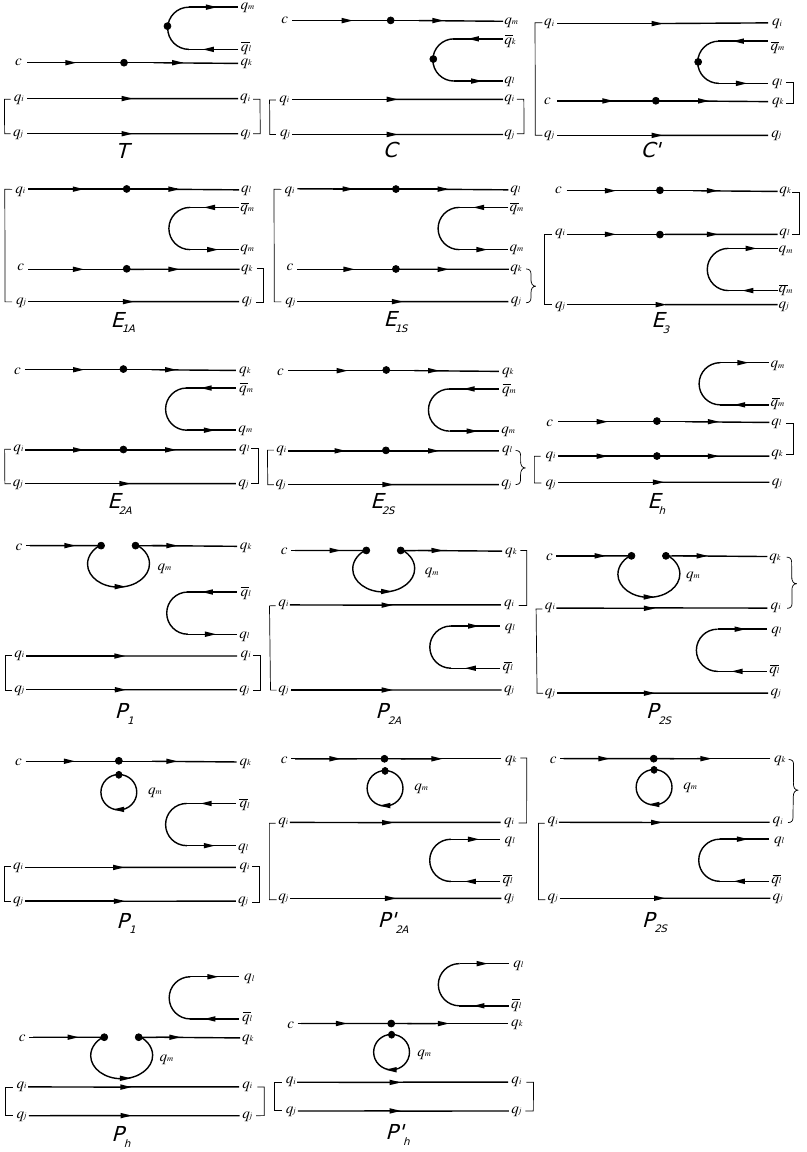}
\caption{Topological diagrams contributing to ${\cal B}_c(\bar 3)\to {\cal B}(8)P(8+1)$ decays. The symbols $]$ and $\}$ denote antisymmetric and symmetric two quark states, respectively. 
}
\label{Fig:TopDiag}
\end{figure}

In terms of the octet baryon wave functions given in Eq. (\ref{eq:wf8}), the relevant topological diagrams for the decays  of antitriplet charmed baryons ${\cal B}_c(\bar 3)\to {\cal B}(8)P(8+1)$ are depicted in Fig. \ref{Fig:TopDiag}: the external $W$-emission, $T$; the internal $W$-emission $C$; the inner $W$-emission $C'$; $W$-exchange diagrams $E_{1A}$, $E_{1S}$, $E_{2A}$, $E_{2S}$, 
$E_3$ and the hairpin diagram $E_h$. Since there are two possible penguin contractions, we will have penguin diagrams $P_1$, $P_{2A}$, $P_{2S}$ as well as $P'_1$, $P'_{2A}$, $P'_{2S}$. The topologies $P_h$ and $P'_h$ are hairpin penguin diagrams. 
The decay amplitudes of ${\cal B}_c(\bar 3)\to {\cal B}(8)P(8+1)$ in the TDA have the expressions \cite{Zhong:2024qqs,Zhong:2024zme}: \begin{equation}
\label{Eq:TDAamp}
\begin{aligned}
\mathcal{A}_{\rm T D A}%\\
=
& \quad T ({\mathcal{B}}_c)^{i j} H_l^{k m}\left({\mathcal{B}}_8\right)_{i j k} (M)_m^l\\
& +C (\mathcal{B}_c)^{i j} H_k^{m l}\left({\mathcal{B}}_8\right)_{i j l} (M)_m^k
 + C' (\mathcal{B}_c)^{i j} H_m^{k l}\left({\mathcal{B}}_8\right)_{klj} (M)_i^m \\
& +E_{1A} (\mathcal{B}_c)^{i j} H_i^{k l}\left({\mathcal{B}}_8\right)_{jkm} (M)_l^m 
 + E_{1S} (\mathcal{B}_c)^{i j} H_i^{k l}(M)_l^m \left[\left({\mathcal{B}}_8\right)_{jmk} 
+\left({\mathcal{B}}_8\right)_{kmj} \right] \\
& +E_{2A} (\mathcal{B}_c)^{i j} H_i^{k l}\left({\mathcal{B}}_8\right)_{jlm} (M)_k^m   + E_{2S} (\mathcal{B}_c)^{i j} H_i^{k l} (M)_k^m\left[\left({\mathcal{B}}_8\right)_{jml}
+ \left({\mathcal{B}}_8\right)_{lmj} \right] \\
&  +E_{3} (\mathcal{B}_c)^{i j} H_i^{k l}\left({\mathcal{B}}_8\right)_{klm} (M)_j^m 
  +E_{h} (\mathcal{B}_c)^{i j} H_i^{k l}\left({\mathcal{B}}_8\right)_{klj} (M)_m^m \\
&  +P_h (\mathcal{B}_c)^{i j} H_m^{m k}\left({\mathcal{B}}_8\right)_{ijk} (M)_l^l 
  +P_{1} (\mathcal{B}_c)^{i j} H_m^{m k}\left({\mathcal{B}}_8\right)_{ijl} (M)_k^l \\  
&  +P_{2A} (\mathcal{B}_c)^{i j} H_m^{m k}\left({\mathcal{B}}_8\right)_{kil} (M)_j^l 
  +P_{2S} (\mathcal{B}_c)^{i j} H_m^{m k}(M)_j^l \left[\left({\mathcal{B}}_8\right)_{kli}+
  \left({\mathcal{B}}_8\right)_{ilk}\right] \\ 
&  +P'_h (\mathcal{B}_c)^{i j} H_m^{k m}\left({\mathcal{B}}_8\right)_{ijk} (M)_l^l 
  +P'_{1} (\mathcal{B}_c)^{i j} H_m^{k m}\left({\mathcal{B}}_8\right)_{ijl} (M)_k^l \\  
&  +P'_{2A} (\mathcal{B}_c)^{i j} H_m^{k m}\left({\mathcal{B}}_8\right)_{kil} (M)_j^l 
  +P'_{2S} (\mathcal{B}_c)^{i j} H_m^{k m}(M)_j^l \left[\left({\mathcal{B}}_8\right)_{kli}+
  \left({\mathcal{B}}_8\right)_{ilk}\right],   
\end{aligned}
\end{equation}
where $(\mathcal{B}_c)^{ij}$ is an antisymmetric baryon matrix standing for antitriplet charmed baryons and  $M^{i}_{j}$ represent nonet mesons, 
\begin{equation}
(\mathcal{B}_c)^{ij}=\left(\begin{array}{ccc}
 0 & \Lambda_c^+ & \Xi_c^+ \\
-\Lambda_c^+& 0 & \Xi_c^0 \\
-\Xi_c^+& -\Xi_c^0 & 0 
\end{array}\right), \qquad  M^{i}_{j}=\left(\begin{array}{ccc}
\frac{\pi^0+ \eta_{q}}{\sqrt{2}} & \pi^+ & K^+ \\
\pi^- & \frac{-\pi^0+ \eta_q}{\sqrt{2}} & K^0 \\
K^- & \overline{K}^0 & \eta_{s}
\end{array}\right),
\end{equation}
with 
\begin{equation}
\label{eq:eta81qs}
    \eta_{8}=\sqrt{\frac{1}{3}}\eta_{q}-\sqrt{\frac{2}{3}}\eta_{s},\quad
    \eta_{1}=\sqrt{\frac{2}{3}}\eta_{q}+\sqrt{\frac{1}{3}}\eta_{s}.
\end{equation}
The physical states $\eta$ and $\eta'$ are given by
\begin{equation}
\left(\begin{array}{c}
	\eta \\
	\eta'
\end{array}\right)=\left(\begin{array}{cc}
	\cos \phi & -\sin \phi\\
	\sin \phi & \cos \phi
\end{array}\right)\left(\begin{array}{l}
	\eta_{q} \\
	\eta_{s}
\end{array}\right)
=\left(\begin{array}{cc}
\cos \theta & -\sin \theta \\
\sin \theta & \cos \theta
\end{array}\right)\left(\begin{array}{c}
\eta_8 \\
\eta_1
\end{array}\right),
\end{equation}
where the mixing angles $\theta$ and $\phi$ are related through the relation $\theta=\phi-\arctan^{-1}\sqrt{2}$. 

The quark content of the octet baryons $\B(8)$ can be read from the subscript $ijk$ of the baryon tensor matrix
$(\mathcal{B}_8)_{i j k} = \epsilon_{ijl} (\mathcal{B}_8)^{l}_{k}$ with
\begin{equation}
(\mathcal{B}_8)^i_j=\left(\begin{array}{ccc}
\frac{1}{\sqrt{6}}\Lambda + \frac{1}{\sqrt{2}}\Sigma^0 & \Sigma^+ & p \\
\Sigma^- & \frac{1}{\sqrt{6}}\Lambda - \frac{1}{\sqrt{2}}\Sigma^0 & n \\
\Xi^-& \Xi^0& -\sqrt{\frac23}\Lambda \end{array}\right), \\
\end{equation}
The tensor coefficient $H^{kl}_m$ related to the CKM matrix elements appears in the standard model Hamiltonian with $H^{kl}_m(\bar q_k c)(\bar q_l q^m)$. The contraction of the two indices of $H^{kl}_m$, namely, $H^{m l}_m$, is induced in the penguin diagrams $P_1, P_{2A}, P_{2S}$ and $P_h$, while $H^{k l}_l$ in the penguin diagrams $P'_1, P'_{2A}, P'_{2S}$ and $P'_h$. 

In the diagrams $T$ and $C$, the two spectator quarks $q_i$ and $q_j$ are antisymmetric in flavor. 
Notice that the final-state quarks $q_l$ and $q_k$  in the topological diagrams $C'$, $E_3$ and $E_h$  also must be antisymmetric in flavor owing to the K\"orner-Pati-Woo (KPW) theorem which states that the quark pair in a baryon produced by weak interactions is required to be antisymmetric in flavor in the SU(3) limit \cite{Korner:1970xq}. Likewise, for $E_{1A,1S}$ and $E_{2A,2S}$, the KPW theorem also leads to 
%\cite{Geng:2018rse} \footnote{
%It should be stressed that the KPW theorem alone is not adequate to lead to Eq. (\ref{eq:E12A}).}
\begin{equation}
\label{eq:E12A}
    E_{2A}=-E_{1A}, \qquad E_{2S}=-E_{1S}.
\end{equation} 
As a result, the number of independent  topological tree diagrams depicted in Fig. \ref{Fig:TopDiag} and the TDA tree amplitudes in Eq. (\ref{Eq:TDAamp}) is 7. 

Working out Eq. (\ref{Eq:TDAamp}) for ${\cal B}_c(\bar 3)\to {\cal B}(8)P(8+1)$ decays,  the obtained TDA decay amplitudes are listed in Tables I and II of Ref. \cite{Zhong:2024qqs} where the penguin contraction 
amplitudes have been neglected.
Among the 7 TDA tree amplitudes given in Eq. (\ref{Eq:TDAamp}), there still exist 2 redundant degrees of freedom through the redefinitions \cite{Chau:1995gk}:
\begin{eqnarray}
\label{eq:tildeTDA}
&& \tilde T=T-E_{1S}, \quad \tilde C=C+E_{1S},\quad \tilde C'=C'-2E_{1S}, \nonumber\\   
%&& \tilde E_2=E_{2A}+E_{2S}+E_3, \quad \tilde E_3=E_{1S}+E_{2S},   \nonumber \\
&& \tilde E_1=E_{1A}+E_{1S}-E_3, \quad \tilde E_h=E_h+2E_{1S}. 
\end{eqnarray}
As a result, among the seven topological tree amplitudes $T$, $C$, $C'$,  $E_{1A}$,  $E_h$, $E_{1S}$ and $E_3$, the last two are redundant degrees of freedom and can be omitted through redefinitions. 
%\footnote{It was claimed in a recent work \cite{Sun:2024mmk} that the relations of $T=C$ and $E_1=E_2$ (or $T_1=T_2$ and $T_4=T_5$ in the notation of Ref. \cite{Sun:2024mmk}) can be employed to reduce the number
%of independent degrees of freedom from 7 to 5. Since the topologies of the color-allowed $T$ and the color-suppressed $C$ are different, the relation of $T=C$ cannot hold and likewise for the relation $E_1=E_2$.}
It is clear that the minimum set of the topological tree amplitudes in the TDA is 5. This is in agreement with the number of tensor invariants found in the IRA \cite{Geng:2023pkr}.

It should be stressed that the redefinition given in Eq. (\ref{eq:tildeTDA}) is not unique. Another redefinition, for example,
\begin{eqnarray}
\label{eq:barTDA}
&& \ov T=T-C'/2, \quad \ov C=C-C'/2, \quad \ov E_{1S}=E_{1S}-C'/2, \nonumber\\   
&& \ov E_{1A}=E_{1A}-E_3+C'/2, \quad \ov E_h=E_h+C',
\end{eqnarray}
is also allowed. 

\subsection{Equivalence of TDA and IRA}
To demonstrate the equivalence between the TDA and IRA, we first follow Ref. \cite{He:2018joe} to write down the general SU(3) invariant decay amplitudes in the IRA: 
\begin{equation}
\begin{aligned}
\label{eq:IRAamp}
\mathcal{A}_{\rm I R Aa}= & \quad a_{1}\left({\B}_c\right)_i\left(H_{\ov 6}\right)_j^{i k}\left({\B}_8\right)_k^j M_l^l
+a_{2}\left({\B}_c\right)_i\left(H_{\ov 6}\right)_j^{i k}\left({\B}_8\right)_k^l M_l^j
+a_{3}\left({\B}_c\right)_i\left(H_{\ov 6}\right)_j^{i k}\left({\B}_8\right)_l^j M_k^l \\
& +a_{4}\left({\B}_c\right)_i\left(H_{\ov 6}\right)_l^{j k}\left({\B}_8\right)_j^i M_k^l
+a_{5}\left({\B}_c\right)_i\left(H_{\ov 6}\right)_l^{j k}\left({\B}_8\right)_j^l M_k^i
+a_{6}\left({\B}_c\right)_i\left(H_{15}\right)_j^{i k}\left({\B}_8\right)_k^j M_l^l \\
& +a_{7}\left({\B}_c\right)_i\left(H_{15}\right)_j^{i k}\left({\B}_8\right)_k^l M_l^j 
+a_{8}\left({\B}_c\right)_i\left(H_{15}\right)_j^{i k}\left({\B}_8\right)_l^j M_k^l
+a_{9}\left({\B}_c\right)_i\left(H_{15}\right)_l^{j k}\left({\B}_8\right)_j^i M_k^l\\
& +a_{10}\left({\B}_c\right)_i\left(H_{15}\right)_l^{j k}\left({\B}_8\right)_j^l M_k^i
+b_1 \Bc_i\left(H_3\right)^j\Boct_j^i M_l^l
+b_2 \Bc_i\left(H_3\right)^j\Boct_j^l M_l^i \\
& +b_3 \Bc_i\left(H_3\right)^i\Boct_j^l M_l^j 
 +b_4 \Bc_i\left(H_3\right)^l\Boct_j^i M_l^j
%+b_5 \Bc_i\left(H_{{15}^b}\right)_l^{j k}\Boct_j^i M_k^l, \\
\end{aligned}
\end{equation}
where the last four terms related to penguins are taken from Ref. \cite{He:2024pxh}. 
%For the explicit expressions of $\left(H_{\ov 6}\right)_k^{i j}$ and $\left(H_{15}\right)_k^{i j}$, see Ref. \cite{He:2018joe}. 
The tensor coefficients $(H_3)^i$ will be introduced later. 
The first five terms associated with $H_{\ov 6}$ are not totally independent as one of them is redundant through the redefinition.  For example, the following two redefinitions 
\begin{equation}
\label{eq:redef2}
 a_1'=a_1-a_5, \quad   a_2'=a_2+a_5, \quad a_3'=a_3+a_5, \quad a_4'=a_4+a_5,  
\end{equation} 
and \cite{He:2018joe}
\begin{equation}
\label{eq:redef1}
 a_1''=a_1+a_2, \quad   a_2''=a_2-a_3, \quad a_3''=a_3-a_4, \quad a_5''=a_5+a_3.  
\end{equation}
are possible. 
As for the five terms associated with $H_{\overline{15}}$ in Eq. (\ref{eqs:IRAamp}), four of them are prohibited by the KPW theorem together with the pole model, namely, $a_6=a_7=a_8=a_{10}=0$ \cite{Geng:2018rse,Geng:2023pkr}.

By comparing the TDA amplitudes in Tables 1 and 2 of \cite{Zhong:2024qqs} with the IRA amplitudes given in Tables 14-16 of Ref. \cite{He:2018joe}, 
we arrive at the relations
\begin{equation}
\label{eqs:TDAtildetoIRA}
\begin{aligned}
&\Tilde{T}=\frac{1}{2}(-a_{2}+a_{4}+a_{9}),\qquad \Tilde{C}=\frac{1}{2}(a_{2}-a_{4}+a_{9}),\\
&\Tilde{C'}=-a_{2}-a_{5}, \qquad \Tilde{E_{1}}=a_{3}+a_{5}, \qquad \Tilde{E_{h}}=-a_{1}+a_{5}.
\end{aligned}
\end{equation}
Therefore, we have the correspondence
\begin{equation}
\begin{aligned}
\label{eq:tildeTDAIRAprime}
& \Tilde{T}=\frac{1}{2}(-a_{2}'+a_{4}'+a_{9}'),\qquad \Tilde{C}=\frac{1}{2}(a_2'-a_{4}'+a_{9}'),\\
& \Tilde{C'}=-a_{2}', \qquad \Tilde{E_{1}}=a_{3}', \qquad \Tilde{E_{h}}=-a_{1}'.
\end{aligned}
\end{equation}
in terms of the redefinitions given in Eq. (\ref{eq:redef2}). The equivalence of the TDA with IRA is thus established. 

There is another set of the IRA tree amplitudes given in Ref. \cite{Geng:2023pkr}
\begin{equation}
\label{eq:IRAb}
\begin{aligned}
\mathcal{A}_{\rm I R A b}^{\rm tree}= & \quad \tilde{f}^a\left({\B}_c\right)^{ik}\left(H_{\ov 6}\right)_{ij}\left({\B}_8\right)_k^j M_l^l+\tilde{f}^b \left({\B}_c\right)^{ik}\left(H_{\ov 6}\right)_{ij}\left({\B}_8\right)_k^l M_l^j+\tilde{f}^c \left({\B}_c\right)^{ik}\left(H_{\ov 6}\right)_{ij}\left({\B}_8\right)_l^j M_k^l \\
& +\tilde{f}^d \left({\B}_c\right)^{kl}\left(H_{\ov 6}\right)_{ij}\left({\B}_8\right)_k^i M_l^j+\tilde{f}^e\left({\B}_c\right)_j\left(H_{15}\right)_l^{i k}\left({\B}_8\right)_i^j M_k^l.
\end{aligned}
\end{equation}
The equivalence between $\widetilde{\rm TDA}$, IRAa and IRAb leads to the relations:
\begin{equation}
\begin{aligned}
\label{eq:tildeTDAIRA1}
&\Tilde{T}
=\frac{1}{2}( \tilde f^b+\tilde f^e),\qquad
\Tilde{C}
=\frac{1}{2}(-\tilde f^b+\tilde f^e), \\
& \Tilde{C'}=\tilde f^b-\tilde f^d,\qquad
\Tilde{E_{1}}=-\tilde f^c,\qquad
\Tilde{E_{h}}=\tilde f^a,\\
\end{aligned}
\end{equation}
or inversely
\begin{equation}
\begin{aligned}
\label{eq:tildeTDAIRA2}
& \tilde f^a=\tilde E_h, \qquad \tilde f^b=\tilde T-\tilde C, \qquad \tilde f^c=-\tilde E_1, \\
& \tilde f^d=\tilde T-\tilde C-\tilde C', \qquad \tilde f^e=\tilde T+\tilde C.
\end{aligned}
\end{equation}

\subsection{From TDA to IRA}
In Refs. \cite{Zhong:2024qqs,Zhong:2024zme} the equivalence of the TDA with IRA is established by first writing down the TDA and IRA amplitudes of ${\cal B}_c(\bar 3)\to {\cal B}(8)P(8+1)$ decays and then comparing them to sort out their relations. Here we will make a direct transformation from the TDA to IRA and show their equivalence (for a similar study, see Ref. \cite{Wang:2024ztg}).

In terms of  $(\mathcal{B}_c)_{i}$ defined by $\frac12\epsilon_{ijk} (\mathcal{B}_c)^{jk}=(\Xi_c^0,-\Xi_c^+,\Lambda_c^+)$ and $(\mathcal{B}_8)^{i}_{j}$, Eq. (\ref{Eq:TDAamp}) can be recast to 
\begin{equation}
\begin{aligned}
\label{eq:TDAamp2}
\mathcal{A}_{\rm TDA}= & \quad (2T-C'-2E_{1S})\Bc_i \Boct_j^i H_m^{j l} M_l^m + (2C+C'+2E_{1S})\Bc_i \Boct_j^i H_m^{l j} M_l^m \\
& +C'\Bc_i \Boct_j^l (H^{ji}_m -H_m^{i j}) M_l^m 
 +(E_{1A}-E_{1S}-E_3)\Bc_i \Boct_m^j (H_j^{i l}-H_j^{l i}) M_l^m \\
 &   +2E_{1S}\Bc_i \Boct_j^m (H_m^{j l}-H_m^{l j}) M_l^i  -E_h\Bc_i \Boct_j^l (H_l^{i j}-H_l^{j i}) M_m^m\\
& +(E_{1S}-E_{1A}+E_3+2P_1-P_{2A}-P_{2S})\Bc_i \Boct_m^i H_j^{jl} M_l^m  \\
& +(-E_{1S}+E_{1A}-E_3+2P'_1-P'_{2A}-P'_{2S})\Bc_i \Boct_m^i H_j^{l j} M_l^m  \\
& +(-E_3+P_{2A}+P_{2S})\Bc_i \Boct_m^l H_j^{j i} M_l^m  
+(E_3+P'_{2A}+P'_{2S})\Bc_i \Boct_m^l H_j^{i j} M_l^m\\
& +(E_h+2P_h+2P_{2S})\Bc_i \Boct_j^i H_l^{l j} M_m^m  +(-E_h+2P'_h+2P'_{2S})
\Bc_i \Boct_j^i H_l^{j l} M_m^m \\
& -2P_{2S}\Bc_i \Boct_j^l H_m^{m j} M_l^i  -2P'_{2S}\Bc_i \Boct_j^l H_m^{j m} M_l^i,
\end{aligned}
\end{equation}
where use of the relations given in Eq. ({\ref{eq:E12A}) has been made.

Consider the effective Hamiltonian responsible for the $\Delta C=1$ weak transition
\begin{equation}
{\cal H}_{\rm eff}=  {G_F\over \sqrt{2}}\left[\sum_{q_1,q_2}^{d,s} V_{cq_1}^*V_{uq_2}(c_1 O_1^{q_1q_2}+c_2 O_2^{q_1q_2})-\lambda_b\sum_{i=3}^6 c_iO_i\right]+ h.c.,
\end{equation}
where $O_1^{q_1q_2}=(\bar uq_2)(\bar q_1 c)$, $O_2^{q_1q_2}=(\bar q_1q_2)(\bar u c)$ with $(\bar q_1q_2)\equiv \bar q_1\gamma_\mu(1-\gamma_5)q_2$, $\lambda_p\equiv V_{cp}^*V_{up}$, and $O_{3-6}$ are QCD-penguin operators. The weak Hamiltonian can be decomposed as ${\bf 3\otimes\bar 3\otimes3}={\bf 3_p\oplus 3_t\oplus \bar 6\oplus 15}$. 
For the singly-Cabibbo-suppressed processes, we have
\begin{equation}
\begin{aligned}
\label{eq:Hdecom}
{\cal H}_{\rm eff}\propto & \quad \lambda_d\left[c_1 (\bar ud)(\bar dc)+c_2(\bar dd)(\bar u c)\right]+\lambda_s
\left[(c_1 (\bar us)(\bar sc)+c_2(\bar ss)(\bar u c)\right] \\
& ={1\over 2}(\lambda_s-\lambda_d)\Big\{ c_+\left[(\bar us)(\bar sc)+(\bar ss)(\bar uc)-(\bar ud)(\bar dc)
-(\bar dd)(\bar uc) \right]_{\bf 15} \\
& \qquad\qquad\quad~~ +c_-\left[(\bar us)(\bar sc)-(\bar ss)(\bar uc)-(\bar ud)(\bar dc)
+(\bar dd)(\bar uc) \right]_{\bf \bar{6}} \Big\}\\
& \quad -{\lambda_b\over 4} \Big\{ c_+\left[(\bar us)(\bar sc)+(\bar ss)(\bar uc)+(\bar ud)(\bar dc)
+(\bar dd)(\bar uc) -2(\bar uu)(\bar uc)\right]_{\bf 15^b} \\
& \quad~ +c_+(O_{3t}+O_{3p})_{\bf 3+}+ 2c_- (O_{3t}-O_{3p})_{\bf 3-} \Big\},
\end{aligned}
\end{equation}
with 
\begin{equation}
\begin{aligned}
& O_{3t}\equiv (\bar u u)(\bar u c)+(\bar u d)(\bar d c)+(\bar u s)(\bar s c)=\sum_{q=u,d,s} (\bar u q)(\bar q c), \\
& O_{3p}\equiv (\bar u u)(\bar u c)+(\bar d d)(\bar u c)+(\bar s s)(\bar u c)=\sum_{q=u,d,s}(\bar qq)(\bar u c),
\end{aligned}
\end{equation}
where $c_\pm\equiv (c_1\pm c_2)/2$  and use of the CKM unitary relation $\lambda_d+\lambda_s+\lambda_b=0$ has been made. Recall that two of the QCD-penguin operators given by \cite{Buchalla:1995vs}
\begin{equation}
O_{3}= \sum_{q=u,d,s} (\bar u_\alpha c_\alpha)(\bar q_\beta q_\beta), \qquad
O_{4}= \sum_{q=u,d,s} (\bar u_\alpha c_\beta)(\bar q_\beta q_\alpha), 
\end{equation}
transform as ${\bf 3_p}$ and ${\bf 3_t}$, respectively.

The Hamiltonian for the current-current operator also can be rewritten as
\begin{equation}
{\cal H}_{\rm eff}\propto \sum_{\lambda=\pm}c_\lambda(H_\lambda)^{ij}_k (\bar q_i c)(\bar q_j q^k)
\end{equation} 
with 
\begin{equation}
\begin{aligned}
(H_+)^{ij}_k &\equiv H^{ij}_k+H^{ji}_k=(H_{15})^{ij}_k+(H_{3+})^i\delta^j_k+(H_{3+})^j\delta^i_k, \\
(H_-)^{ij}_k &\equiv H^{ij}_k-H^{ji}_k= (H_{\ov 6})^{ij}_k+2(H_{3-})^i\delta^j_k-2(H_{3-})^j\delta^i_k, \\
\end{aligned}
\end{equation}
where $H_{3\pm}=H_{3t}\pm H_{3p}$. 
The coefficient tensor $H$ then reads \cite{Grossman:2012ry}
\begin{equation}
\label{eq:Hijk}
H_k^{ij}=\frac12\left[
(H_{15})_k^{ij} +(H_{\ov 6})_k^{ij} \right] + \delta^j_k\left(\frac32 (H_{3p})^i-\frac12 (H_{3t})^i\right)+
 \delta^i_k\left(\frac32 (H_{3t})^j-\frac12 (H_{3p})^j\right).
\end{equation} 
Note that our convention for $H^{ij}_k(\bar q_i c)(\bar q_j q^k)$
is different from that given in Ref. \cite{Grossman:2012ry}  in regard to the order of $ij$ indices. 
It is easily seen  that  $H^{k i}_k=4(H_{3t})^i$ and $H^{i k}_k=4(H_{3p})^i$.  The explicit expressions of the coefficient tensor can be read from Eq. (\ref{eq:Hdecom}) as (see also Ref. \cite{Xing:2024nvg})
\begin{equation}
\begin{aligned}
\label{eq:Htensor}
& (H_{\ov 6})^{31}_3=-(H_{\ov 6})^{13}_3=(H_{\ov 6})^{12}_2=-(H_{\ov 6})^{21}_2=\frac12(\lambda_s-\lambda_d), \\
& (H_{15})^{31}_3= (H_{15})^{13}_3=-(H_{15})^{12}_2=-(H_{15})^{21}_2=\frac12(\lambda_s-\lambda_d), \\
& (H_{{15}^b})^{31}_3= (H_{{15}^b})^{13}_3=(H_{{15}^b})^{12}_2=(H_{{15}^b})^{21}_2=-\frac14 \lambda_b, \\
& (H_{{15}^b})^{11}_1=\frac12\lambda_b, \quad  (H_{3_-})^1=(H_{3_+})^1=-\frac14 \lambda_b,\quad
 (H_{3t})^1=-\frac14\lambda_b,\quad (H_{3p})^1=0.  \\
\end{aligned}
\end{equation}
For later convenience, we have followed Ref. \cite{He:2024pxh} to lump the components of $(H_{15})^{ij}_k$  proportional to $\lambda_b$ into $(H_{{15}^b})^{ij}_k$.

For Cabibbo-favored and doubly-Cabibbo-suppressed decays we have
\begin{equation}
\begin{aligned}
& (H_{\ov 6})^{31}_2=-(H_{\ov 6})^{13}_2= (H_{15})^{31}_2= (H_{15})^{13}_2= V_{cs}^*V_{ud}, \\
& (H_{\ov 6})^{21}_3=-(H_{\ov 6})^{12}_3= (H_{15})^{21}_3= (H_{15})^{12}_3= V_{cd}^*V_{us}. \\
\end{aligned}
\end{equation}

Substituting the relation (\ref{eq:Hijk})
into Eq. (\ref{eq:TDAamp2}) yields
\begin{equation}
\mathcal{A}_{\rm TDA}=\mathcal{A}_{\rm TDA}^{\rm tree}+\mathcal{A}_{\rm TDA}^{\lambda_b}
\end{equation}
with
\begin{equation}
\begin{aligned}
\label{eq:TDAtree}
\mathcal{A}_{\rm TDA}^{\rm tree}= & \quad (T+C)\Bc_i\left(H_{15}\right)_m^{j l}\Boct_j^i M_l^m 
 - E_h\Bc_i \left(H_{\ov 6}\right)_l^{i j}\Boct_j^l M_m^m \\
& + (T-C-C'-2E_{1S})\Bc_i\left(H_{\ov 6}\right)_m^{j l}\Boct_j^i M_l^m 
 -C'\Bc_i \left(H_{\ov 6}\right)_m^{i j}\Boct_j^l M_l^m \\
& +(E_{1A}-E_{1S}-E_3)\Bc_i \left(H_{\ov 6}\right)_j^{i l}\Boct_m^j M_l^m +2 E_{1S}\Bc_i \left(H_{\ov 6}\right)_m^{j l}\Boct_j^m M_l^i  \\
\end{aligned}
\end{equation}
and 
\begin{equation}
\begin{aligned}
\label{eq:TDAlambdab}
\mathcal{A}_{\rm TDA}^{\lambda_b}=  & \quad b_1 \Bc_i\left(H_{3p}\right)^j\Boct_j^i M_l^l
+b_2 \Bc_i\left(H_{3p}\right)^j\Boct_j^l M_l^i
+b_3 \Bc_i\left(H_{3p}\right)^i\Boct_j^l M_l^j \\
& +b_4 \Bc_i\left(H_{3p}\right)^l\Boct_j^i M_l^j + b_5 \Bc_i\left(H_{3t}\right)^j\Boct_j^i M_l^l
+b_6 \Bc_i\left(H_{3t}\right)^j\Boct_j^l M_l^i\\
& +b_7 \Bc_i\left(H_{3t}\right)^i\Boct_j^l M_l^j +b_8 \Bc_i\left(H_{3t}\right)^l\Boct_j^i M_l^j
 +b_9 \Bc_i\left(H_{{15}^b}\right)^{jk}_l\Boct_j^i M_k^l, \\
\end{aligned}
\end{equation}
where
\begin{equation}
\begin{aligned}
\label{eqs:IRAamp}
& b_1=3 T-C-2 (C'+2E_{1S})-2 E_h+8P'_h+8P'_{2S}, \\
& b_2= 2 C'+4E_{1S}-8P'_{2S},    \\
& b_3=-2 C'+2(E_{1A}-E_{1S}+E_3)+4P'_{2A}+4P'_{2S},\\
& b_4= - T+3 C+2(C'+E_{1S})+2(E_{1A}-E_3)+8P'_1-4P'_{2A}-4P'_{2S}. \\
& b_5=- T+3 C+2 (C'+2E_{1S})+2 E_h+8P_h+8P_{2S}, \\
& b_6= -2 C'-4E_{1S}-8P_{2S}, \\
& b_7= 2 C'-2(E_{1A}-E_{1S}+E_3)+4P_{2A}+4P_{2S},\\
& b_8=  3 T- C-2(C'+E_{1S})-2(E_{1A}-E_3)+8P_1-4P_{2A}-4P_{2S}, \\
& b_9=T+C. \\
\end{aligned}
\end{equation}
The $\mathcal{A}_{\rm TDA}^{\lambda_b}$ involves penguin-like contributions induced from the current-current
tree operators $O_{1,2}$ or penguin contributions arising from QCD penguin operators. 

From the expression of  $\mathcal{A}_{\rm TDA}^{\rm tree}$, we see  that there is only one term related to $H_{15}$.  Comparing it  with Eq. (\ref{eq:IRAamp}) yields
\begin{equation}
\begin{aligned}
& a_1=- E_h, \qquad a_2=-C', \qquad a_3= E_{1A}-E_{1S}-E_3, \\
& a_4= T-C-C'-2E_{1S}, \qquad a_5=2E_{1S}, \qquad a_9=T+C
\end{aligned}
\end{equation}
and hence
\begin{equation}
\begin{aligned}
& a_1'=- \tilde{E}_h, \qquad a_2'= -\tilde C', \qquad a_3'= \tilde{E}_1,\\
& a_4'= \tilde T-\tilde C-\tilde C',
\qquad a_9=\tilde T+\tilde C, \\
\end{aligned}
\end{equation}
consistent with Eq. (34) of \cite{Zhong:2024qqs}}, where $a'_i$ are defined in Eq. (\ref{eq:redef2}).

From the expressions of $(H_{3p})^i$, $(H_{3t})^i$ and $(H_{15^b})^{ij}_k$ given in Eq. (\ref{eq:Htensor}),  it is obvious that
$\mathcal{A}_{\rm TDA}^{\lambda_b}$  is proportional to $\lambda_b$ and can be recast to
\begin{equation}
\begin{aligned}
\label{eq:TDAlambdab2}
\mathcal{A}_{\rm TDA}^{\lambda_b}=  & \quad \tilde b_1 \Bc_i\left(H_3\right)^j\Boct_j^i M_l^l
+\tilde b_2 \Bc_i\left(H_3\right)^j\Boct_j^l M_l^i
+\tilde b_3 \Bc_i\left(H_3\right)^i\Boct_j^l M_l^j \\
& +\tilde b_4 \Bc_i\left(H_3\right)^l\Boct_j^i M_l^j 
 +\tilde b_5 \Bc_i\left(H_{{15}^b}\right)^{jk}_l\Boct_j^i M_k^l, \\
\end{aligned}
\end{equation}
where the matrix element $H_3$ is normalized to  $(1,0,0)$ and  \footnote{ The expressions in Eq. (\ref{eq:bprime}) are obtained using $(H_{3_-})^1=(H_{3_+})^1=-\frac14 \lambda_b$ (see Eq. (\ref{eq:Htensor})). In the presence of QCD penguin operators $O_3$ and $O_4$, the tensor matrix elements are modified to $(H_{3_-})^1=-\frac14 \lambda_b[1+(c_3-c_4)/c_-]$ and $(H_{3_+})^1=-\frac14 \lambda_b[1+(2c_3+2c_4)/c_+]$. We wish to thank Chia-Wei Liu for pointing out this to us.}
\begin{equation}
\tilde b_1=-\frac14 b_5, \qquad \tilde b_2=-\frac14 b_6, \qquad \tilde b_3=-\frac14 b_7, \qquad \tilde b_4=-\frac14 b_8, \qquad \tilde b_5=b_9.
\end{equation}
which can be expressed in terms of tilde quantities
\begin{equation}
\begin{aligned}
\label{eq:bprime}
& \tilde b_1=\frac14(\tilde T-3\tilde  C)-\frac12 \tilde C'-\frac12 \tilde E_h-2\tilde P_h-2\tilde P_{2S}, \\
& \tilde b_2= \frac12 \tilde C' +2\tilde P_{2S}, \\
& \tilde b_3= -\frac12 \tilde C'+\frac12\tilde E_{1A}-\tilde P_{2A}-\tilde P_{2S},\\
& \tilde b_4=  -\frac14(3 \tilde T-\tilde C)+\frac12\tilde C'+\frac12\tilde E_{1}-2\tilde P_1+\tilde P_{2A}+\tilde P_{2S}, \\
&  \tilde b_5=\tilde T+\tilde C,  \\
\end{aligned}
\end{equation} 
with 
\begin{equation}
\begin{aligned}
\tilde P_1 & = P_1+E_{1S}-\frac12 E_3, \qquad   \tilde P_h  = P_h - E_{1S}, \\
\tilde P_{2S} & = P_{2S}+E_{1S} , \qquad  \qquad ~\tilde P_{2A}  = P_{2A}+E_{1S}-E_3, \\
\tilde P'_1 & = P'_1-E_{1S}+\frac12 E_3, \qquad  \tilde P'_h  = P'_h + E_{1S}, \\
\tilde P'_{2S} & = P'_{2S}-E_{1S} , \qquad  \qquad ~\tilde P'_{2A}  = P'_{2A}-E_{1S}+E_3.  
\end{aligned}
\end{equation}

Comparing $\mathcal{A}_{\rm TDA}^{\lambda_b}$ with the $\lambda_b$ terms in the IRAb \cite{He:2024pxh}
\begin{equation}
\begin{aligned}
\label{eq:IRAHe}
\mathcal{A}_{\rm IRAb}^{\lambda_b}=  & \quad \tilde{f}^a_3 \Bc_i\left(H_3\right)^j\Boct_j^i M_k^k
+\tilde{f}^b_3 \Bc_k\left(H_3\right)^i\Boct_i^l M_l^k 
 +\tilde{f}^c_3 \Bc_i\left(H_3\right)^i\Boct_j^l M_l^j \\
&  +\tilde{f}^d_3 \Bc_i\left(H_3\right)^l\Boct_j^i M_l^j +\tilde{f}^e \Bc_i\left(H_{{15}^b}\right)^{jk}_l\Boct_j^i M_k^l,
\\
\end{aligned}
\end{equation}
we have 
\begin{equation}
\tilde b_1=\tilde{f}_3^a, \qquad \tilde b_2=\tilde{f}^b_3, \qquad \tilde b_3=\tilde{f}^c_3, \qquad \tilde b_4=\tilde{f}^d_3, \qquad
\tilde b_5=\tilde{f}^e.
\end{equation}
In the TDA, the relations of $\tilde b_1,\cdots, \tilde b_4$ with the penguin diagrams depicted in Fig. \ref{Fig:TopDiag} 
are explicitly shown in Eq. (\ref{eq:bprime}).
Writing $\mathcal{A}_{\rm IRAb}=\mathcal{A}_{\rm IRAb}^{\rm tree}+\mathcal{A}_{\rm IRAb}^{\lambda_b}$,
we see from Eqs. (\ref{eq:IRAb}) and (\ref{eq:IRAHe}) that there are totally 9 independent amplitudes, 5 in  
$\mathcal{A}_{\rm IRAb}^{\rm tree}$ and 4 in $\mathcal{A}_{\rm IRAb}^{\lambda_b}$. Likewise, there are 5 independent amplitudes in  
$\mathcal{A}_{\rm TDA}^{\rm tree}$ and 4 in $\mathcal{A}_{\rm TDA}^{\lambda_b}$.

\section{Numerical Analysis and Results}
\label{sec:Num}
Neglecting the penguin contraction terms,  there exist 5 independent tilde TDA amplitudes given in Eq. (\ref{eq:tildeTDA}). Hence, we have totally 19 unknown parameters to describe the magnitudes and the phases of the respective $S$- and $P$-wave amplitudes, namely, 
\begin{eqnarray}
\label{eq:19parameters}
&& |\tilde T|_Se^{i\delta_S^{\tilde T}}, \quad |\tilde C|_Se^{i\delta_S^{\tilde C}}, \quad |\tilde C'|_Se^{i\delta_S^{\tilde C'}},  \quad |\tilde E_{1}|_Se^{i\delta_S^{\tilde E_{1}}}, \quad |\tilde E_h|_Se^{i\delta_S^{\tilde E_h}},\nonumber \\
&& |\tilde T|_Pe^{i\delta_P^{\tilde T}}, \quad |\tilde C|_Pe^{i\delta_P^{\tilde C}}, \quad |\tilde C'|_Pe^{i\delta_P^{\tilde C'}}, \quad |\tilde E_{1}|_Pe^{i\delta_P^{\tilde E_{1}}}, \quad |\tilde E_h|_Pe^{i\delta_P^{\tilde E_h}},
\end{eqnarray}
collectively denoted by $|X_i|_Se^{i\delta^{X_i}_S}$ and $|X_i|_Pe^{i\delta^{X_i}_P}$,
where the subscripts $S$ and $P$ denote the $S$- and $P$-wave components of each TDA amplitude. 
Since there is an overall phase which can be omitted, we shall set $\delta_S^{\tilde T}=0$.  Likewise, for the tidle IRA amplitudes given in Eq. (\ref{eq:IRAb}), we also have \footnote{Note that $S$- and $P$-wave amplitudes in Ref. \cite{Geng:2023pkr}  were denoted by $\tilde f^i$ and $\tilde g^i$ with $i=a,b,c,d,e$, respectively.} 
\begin{eqnarray}
\label{eq:IRA19parameters}
&& |\tilde f^a|_Se^{i\delta_S^{\tilde f^a}}, \quad |\tilde f^b|_Se^{i\delta_S^{\tilde f^b}}, \quad |\tilde f^c|_Se^{i\delta_S^{\tilde f^c}},  \quad |\tilde f^d|_Se^{i\delta_S^{\tilde f^d}}, \quad |\tilde f^e|_Se^{i\delta_S^{\tilde f^e}},\nonumber \\
&& |\tilde f^a|_Pe^{i\delta_P^{\tilde f^a}}, \quad |\tilde f^b|_Pe^{i\delta_P^{\tilde f^b}}, \quad |\tilde f^c|_Pe^{i\delta_P^{\tilde f^c}},  \quad |\tilde f^d|_Pe^{i\delta_P^{\tilde f^d}}, \quad |\tilde f^e|_Pe^{i\delta_P^{\tilde f^e}}.
\end{eqnarray}
We shall aslo set $\delta_S^{\tilde f^a}=0$. Of course, physics is independent of which phase is removed. 
Hence, in both cases we are left with 19 parameters.

In terms of the $S$- and $P$-wave amplitudes given in Eq. (\ref{eq:A&B})
and their corresponding phases $\delta_S$ and $\delta_P$, the decay rate and Lee-Yang parameters are given in Eqs. (\ref{eq:Gamma}) and (\ref{eq:decayparameter}), respectively.
 It is clear that the magnitudes of $S$- and $P$-wave amplitudes can be determined from $\Gamma$ and $\gamma$. As for the phase shift between $S$- and $P$-wave amplitudes, it is naively expected that  $\delta_P - \delta_S = \arctan({\beta}/{\alpha})$ (see, for example, Ref. \cite{BESIII:XiK}). However, this is  is somewhat inconvenient
as the range of this solution is limited to $(-\frac{\pi}{2}, \frac{\pi}{2})$, which does not 
fully cover the phase-shift space. Hence, it often leads to an ambiguity of $\pm\pi$ rad.
A suitable one is \cite{Zhong:2024qqs}
\begin{equation}
\label{eq:phase}
\delta_P - \delta_S = 2 \arctan \frac{\beta}{\sqrt{\alpha^2+\beta^2}+\alpha}.
\end{equation}
This  naturally covers the correct solution space without imposing manual modification. 

As noticed in Ref. \cite{Geng:2023pkr}, there is a so-called $Z_2$ ambiguity in the determination of  phases, namely, $(\delta_S^{X_i}, \delta_P^{X_i})\to (-\delta_S^{X_i}, -\delta_P^{X_i})$. Since $\beta$ is proportional to $\sin(\delta_P-\delta_S)$, its sign ambiguity can be resolved by the measurement of $\beta$ as just noticed in passing. The sign of the Lee-Yang parameter $\gamma$ depends on the relative magnitude of $S$- and $P$-wave amplitudes. Hence we also need measurements of $\gamma$ to select the solutions for $|A|$ and
$|B|$. Recently, LHCb has performed the first measurements of all the Lee-Yang parameters in $\Lambda_c^+\to\Lambda \pi^+$ and $\Lambda_c^+\to \Lambda K^+$, see Table \ref{tab:LHCb}.

%%%%%%%%%%%%%%%%%%
%----------Scenario C----------
\begin{table}[t]
\caption{Fitted tilde TDA (upper) and IRA (lower) amplitudes in Case I. We have set $\delta^{\tilde{T}}_S=0$   and $\delta^{\tilde{f^a}}_S=0$.} 
\label{tab:ampScenI}
\vspace{-0.1cm}
\begin{center}
\renewcommand\arraystretch{1}
% \resizebox{\textwidth}{!} 
% {
\begin{tabular}
{c| c r r r }
\hline \hline
&$|X_i|_S$&$|X_i|_P$~~~ &$\delta^{X_i}_{S}$~~~~~ &$\delta^{X_i}_{P}$~~~~ \\
&\multicolumn{2}{c}{$(10^{-2}G_{F}~{\rm GeV}^2)$}&\multicolumn{2}{c}{$(\text{in radian})$} \\
\hline 
$\Tilde{T}$&
~$4.25\pm0.11$&$12.43\pm0.31$&
-- ~~~~~~~&$2.40\pm0.04$\\
$\Tilde{C}$&
~$3.08\pm0.50$&$11.57\pm0.96$&
$3.02\pm0.12$&$-0.77\pm0.20$\\
$\Tilde{C'}$&
~$5.39\pm0.39$&$18.79\pm0.88$&
$-0.03\pm0.05$&$2.23\pm0.11$\\
$\Tilde{E_{1}}$&
~$2.90\pm0.19$&$10.22\pm0.49$&
$-2.80\pm0.05$&$1.87\pm0.10$\\
$\Tilde{E_{h}}$&
~$4.06\pm0.53$&$13.82\pm1.93$&
$2.66\pm0.12$&$-1.90\pm0.20$\\
\hline
%20241205
$\tilde{f}^a$&
~$4.10\pm0.52$&$16.18\pm2.35$&
-- ~~~~~~~&$1.72\pm0.12$\\
$\tilde{f}^b$&
~$7.00\pm1.40$&$24.52\pm1.78$&
$-2.78\pm0.11$&$-0.30\pm0.19$\\
$\tilde{f}^c$&
~$2.91\pm0.19$&$10.26\pm0.49$&
$-2.36\pm0.11$&$2.29\pm0.15$\\
$\tilde{f}^d$&
~$1.59\pm1.23$&$7.50\pm4.00$&
$-2.89\pm0.41$&$0.25\pm0.41$\\
$\tilde{f}^e$&
~$1.57\pm1.44$&$0.71\pm3.02$&
$-2.44\pm0.23$&$-1.64\pm3.66$\\
\hline \hline
\end{tabular}
% }
\end{center}
\end{table}
%%%%%%%%%%%%%

%%%%%%%%%%%%%%%%%%%%%%
\begin{table}[t]
\caption{Same as Table \ref{tab:ampScenI} except in Case II.}
\label{tab:ampScenII}\vspace{-0.1cm}
\begin{center}
\renewcommand\arraystretch{1}
% \resizebox{\textwidth}{!} 
% {
\begin{tabular}%{\textwidth}
{c| c r r r }
\hline \hline
&$|X_i|_S$&$|X_i|_P$~~~ &$\delta^{X_i}_{S}$~~~~~ &$\delta^{X_i}_{P}$~~~~ \\
&\multicolumn{2}{c}{$(10^{-2}G_{F}~{\rm GeV}^2)$}&\multicolumn{2}{c}{$(\text{in radian})$} \\
\hline
$\Tilde{T}$&
~$4.22\pm0.10$&$12.50\pm0.28$&
-- ~~~~~~~&$2.42\pm0.04$\\
$\Tilde{C}$&
~$2.40\pm0.66$&$12.70\pm0.71$&
$2.89\pm0.59$&$-0.57\pm0.15$\\
$\Tilde{C'}$&
~$5.26\pm0.35$&$19.04\pm0.85$&
$-0.02\pm0.05$&$2.32\pm0.11$\\
$\Tilde{E_{1}}$&
~$2.86\pm0.19$&$10.20\pm0.49$&
$-2.80\pm0.05$&$1.83\pm0.09$\\
$\Tilde{E_{h}}$&
~$3.07\pm0.47$&$11.80\pm1.42$&
$2.87\pm0.09$&$-1.76\pm0.19$\\
\hline
$\tilde{f}^a$&
~$3.16\pm0.43$&$10.74\pm1.73$&
-- ~~~~~~~&$1.68\pm0.15$\\
$\tilde{f}^b$&
~$7.52\pm0.31$&$23.27\pm0.69$&
$-2.98\pm0.09$&$-0.56\pm0.10$\\
$\tilde{f}^c$&
~$2.86\pm0.19$&$10.19\pm0.49$&
$-2.51\pm0.09$&$2.13\pm0.12$\\
$\tilde{f}^d$&
~$2.34\pm0.20$&$4.30\pm0.65$&
$3.02\pm0.20$&$-0.75\pm0.39$\\
$\tilde{f}^e$&
~$1.48\pm0.32$&$3.82\pm1.04$&
$-2.04\pm0.17$&$0.60\pm0.15$\\
\hline \hline
\end{tabular}
% }
\end{center}
\end{table}
%%%%%%%%%%%

%%%%%%%%%%%%%%%%%%%%%%%%%%%%%%%
\begin{table}[t]
\caption{The $\chi^2$ values of the TDA and IRA fits in Cases I and II. }
 \label{tab:chi}
\begin{center} 
\begin{tabular}{l c c c  c}
\hline \hline
& ~~~~Case I~ & & ~~~~Case II & \\
\hline
& TDA & IRA & TDA & IRA \\
\hline
$\chi^2$ & 34.31 & 33.21 & 59.17 & 57.08 \\
$\chi^2/d.o.f.$ & 2.29 & 2.21 & 3.11 & 3.00 \\
\hline \hline
\end{tabular}
\end{center} 
\end{table}
%%%%%%%%%%%%%%%%%%%%%%%%%%%%%%%%%%%

All the currently available 38 data are collected in the Appendix, see Table \ref{tab:expandave}. For our purpose, we shall consider two different fits: Case I without the recent Belle data on $\Xi_c^0\to \Xi^0 \pi^0, \Xi^o\eta^{(')}$  \cite{Belle-II:2024jql} and Case II with all the data included. 
Fitted tilde TDA and IRA amplitudes collectively denoted by $\tilde{X}_i$ are shown in Tables \ref{tab:ampScenI}
and \ref{tab:ampScenII} for Cases I and II, respectively.  
We have set $\delta^{\tilde{T}}_S=0$
and $\delta^{\tilde{f^a}}_S=0$. The $\chi^2$ values of the TDA and IRA fits in Cases I and II are shown in Table \ref{tab:chi}. It is clear that the $\chi^2$ value per degree of freedom is around 2.2 in Case I, but slightly higher in Case II. 
The fit results of branching fractions, decay parameters, the magnitudes of $S$- and $P$-wave amplitudes and their phase shifts within the framework of  the tilde TDA  and tilde IRA are shown in Tables \ref{tab:resultScenI}-\ref{tab:fitother2ScenII} for Cabibbo-favored (CF), singly Cabibbo-suppressed (SCS) and doubly Cabibbo-suppressed (DCS) decay modes.
\vskip 0.2cm
In the following we discuss their implications.

%%%%%%%%%%%%%%
\begin{table*}[h]
\caption{
The fit results based on the tilde TDA (upper) and tilde IRA (lower) in Case I. Experimental data from Belle \cite{Belle-II:2024jql}  with an asterisk are not included in the fit.
$S$- and $P$-wave amplitudes are in units of $10^{-2}G_F~{\rm GeV}^2$ and $\delta_P-\delta_S$ in radian. }
\label{tab:resultScenI}
\centering
\resizebox{\textwidth}{!} 
{
\begin{tabular}
{ l |c rrr rrr |  c c  c }
\hline
\hline
\multirow{2}{*}{Channel}
&\multirow{2}{*}{$10^{2}\mathcal{B}$}
&\multirow{2}{*}{$\alpha$}~~~~~~~
&\multirow{2}{*}{$\beta$}~~~~~~~
&\multirow{2}{*}{$\gamma$}~~~~~~
&\multirow{2}{*}{$|A|$}~~~~
&\multirow{2}{*}{$|B|$}~~~~ 
&\multirow{2}{*}{$\delta_P-\delta_S$}~~~ 
&\multirow{2}{*}{$10^{2}\mathcal{B}_\text{exp}$}
&\multirow{2}{*}{$\alpha_\text{exp}$}
&$\beta_\text{exp}$\\
&&&&&&&&&&$\gamma_\text{exp}$\\
\hline
\multirow{2}{*}
{$\Lambda_c^+\to\Lambda^0\pi^+$}&$1.30\pm0.05$&$-0.76\pm0.01$&$0.39\pm0.02$&$0.51\pm0.01$&$5.57\pm0.10$&$9.24\pm0.20$&$2.67\pm0.02$&\multirow{2}{*}{$1.29\pm{0.05}$}&\multirow{2}{*}{$-0.762\pm0.006$}&$0.368\pm0.021$\\&$1.30\pm0.05$&$-0.76\pm0.01$&$0.39\pm0.02$&$0.51\pm0.01$&$5.57\pm0.10$&$9.23\pm0.20$&$2.67\pm0.02$&&&$0.502\pm0.017$\\
\multirow{2}{*}{$\Lambda_c^+\to\Sigma^0\pi^+$}&$1.25\pm0.05$&$-0.47\pm0.01$&$0.35\pm0.10$&$-0.81\pm0.05$&$1.94\pm0.23$&$19.20\pm0.48$&$2.50\pm0.14$&\multirow{2}{*}{$1.27\pm{0.06}$}&\multirow{2}{*}{$-0.466\pm0.018$}\\&$1.25\pm0.05$&$-0.47\pm0.01$&$0.36\pm0.10$&$-0.80\pm0.05$&$1.97\pm0.23$&$19.14\pm0.47$&$2.48\pm0.13$\\
\multirow{2}{*}{$\Lambda_c^+\to\Sigma^+ \pi^0$}&$1.26\pm0.05$&$-0.47\pm0.01$&$0.35\pm0.10$&$-0.81\pm0.05$&$1.94\pm0.23$&$19.20\pm0.48$&$2.50\pm0.14$&\multirow{2}{*}{$1.24\pm0.09$}&\multirow{2}{*}{$-0.484\pm0.027$}\\&$1.26\pm0.05$&$-0.47\pm0.01$&$0.36\pm0.10$&$-0.81\pm0.05$&$1.97\pm0.23$&$19.14\pm0.47$&$2.48\pm0.13$\\
\multirow{2}{*}{$\Lambda_c^+\to\Sigma^+ \eta$}&$0.33\pm0.04$&$-0.92\pm0.04$&$-0.01\pm0.15$&$0.40\pm0.10$&$2.94\pm0.21$&$6.98\pm0.74$&$-3.13\pm0.16$&\multirow{2}{*}{$0.32\pm0.05$}&\multirow{2}{*}{$-0.99\pm0.06$}\\&$0.32\pm0.04$&$-0.92\pm0.04$&$-0.15\pm0.16$&$0.36\pm0.12$&$2.87\pm0.20$&$7.15\pm0.83$&$-2.98\pm0.17$\\
\multirow{2}{*}{$\Lambda_c^+\to\Sigma^+ \eta'$}&$0.39\pm0.07$&$-0.44\pm0.07$&$0.88\pm0.06$&$0.16\pm0.28$&$4.03\pm0.78$&$21.52\pm2.63$&$2.03\pm0.08$&\multirow{2}{*}{$0.41\pm0.08$}&\multirow{2}{*}{$-0.46\pm0.07$}\\&$0.43\pm0.07$&$-0.45\pm0.07$&$0.90\pm0.03$&$-0.02\pm0.30$&$3.91\pm0.80$&$24.80\pm3.36$&$2.03\pm0.08$\\
\multirow{2}{*}{$\Lambda_c^+\to\Xi^0 K^+$}&$0.34\pm0.03$&$-0.04\pm0.12$&$-0.98\pm0.02$&$0.19\pm0.09$&$2.76\pm0.18$&$9.71\pm0.47$&$-1.61\pm0.12$&\multirow{2}{*}{$0.55\pm0.07$}&\multirow{2}{*}{$0.01\pm0.16$}\\&$0.34\pm0.03$&$-0.06\pm0.12$&$-0.98\pm0.02$&$0.19\pm0.09$&$2.77\pm0.18$&$9.75\pm0.46$&$-1.63\pm0.12$\\
\multirow{2}{*}{$\Lambda_c^+\to p K_S$}&$1.56\pm0.06$&$-0.74\pm0.03$&$0.56\pm0.15$&$-0.37\pm0.22$&$4.17\pm0.71$&$15.07\pm1.37$&$2.50\pm0.14$&\multirow{2}{*}{$1.59\pm0.07$}&\multirow{2}{*}{$-0.743\pm0.028$}\\&$1.59\pm0.06$&$-0.74\pm0.03$&$0.44\pm0.56$&$-0.51\pm0.48$&$3.73\pm1.82$&$16.60\pm2.72$&$2.61\pm0.56$\\
\multirow{2}{*}{$\Xi_c^0\to\Xi^- \pi^+$}&$2.97\pm0.09$&$-0.73\pm0.03$&$0.67\pm0.03$&$0.13\pm0.04$&$8.07\pm0.21$&$23.63\pm0.58$&$2.40\pm0.04$&\multirow{2}{*}{$1.80\pm0.52$}&\multirow{2}{*}{$-0.64\pm0.05$}\\&$2.96\pm0.09$&$-0.72\pm0.03$&$0.68\pm0.03$&$0.13\pm0.04$&$8.08\pm0.22$&$23.47\pm0.58$&$2.38\pm0.04$\\
\multirow{2}{*}{$\Xi_c^0\to\Xi^0 \pi^0$}&$0.72\pm0.04$&$-0.64\pm0.07$&$0.77\pm0.06$&$-0.06\pm0.11$&$3.62\pm0.26$&$12.63\pm0.59$&$2.26\pm0.10$&\multirow{2}{*}{$0.69\pm0.16^{*}$}&\multirow{2}{*}{$-0.90\pm0.27^{*}$}\\&$0.71\pm0.04$&$-0.61\pm0.08$&$0.79\pm0.06$&$-0.04\pm0.10$&$3.65\pm0.25$&$12.49\pm0.58$&$2.22\pm0.10$\\
\multirow{2}{*}{$\Xi_c^0\to\Xi^0 \eta$}&$0.26\pm0.04$&$0.23\pm0.15$&$-0.09\pm0.15$&$-0.97\pm0.04$&$0.43\pm0.27$&$12.58\pm1.08$&$-0.36\pm0.56$&\multirow{2}{*}{$0.16\pm0.05^{*}$}\\&$0.23\pm0.04$&$0.23\pm0.15$&$-0.15\pm0.15$&$-0.96\pm0.05$&$0.45\pm0.26$&$11.75\pm0.98$&$-0.57\pm0.49$\\
\multirow{2}{*}{$\Xi_c^0\to\Xi^0 \eta'$}&$0.43\pm0.06$&$-0.70\pm0.06$&$0.71\pm0.06$&$-0.07\pm0.27$&$3.91\pm0.75$&$23.72\pm2.56$&$2.35\pm0.08$&\multirow{2}{*}{$0.12\pm0.04^{*}$}\\&$0.49\pm0.07$&$-0.67\pm0.06$&$0.70\pm0.09$&$-0.25\pm0.28$&$3.73\pm0.77$&$27.18\pm3.33$&$2.33\pm0.09$\\
\multirow{2}{*}{$\Xi_c^+\to\Xi^0\pi^+$}&$0.98\pm0.13$&$-0.88\pm0.08$&$0.31\pm0.10$&$0.36\pm0.13$&$2.95\pm0.23$&$6.67\pm0.84$&$2.80\pm0.13$&\multirow{2}{*}{$1.6\pm0.8$}&\\&$1.00\pm0.13$&$-0.90\pm0.07$&$0.29\pm0.10$&$0.32\pm0.13$&$2.93\pm0.22$&$6.94\pm0.86$&$2.83\pm0.12$\\
\hline
\multirow{2}{*}{$\Lambda_c^+\to\Lambda^0 K^+$}&$0.0635\pm0.0030$&$-0.58\pm0.04$&$0.40\pm0.07$&$-0.71\pm0.04$&$0.57\pm0.04$&$4.41\pm0.11$&$2.53\pm0.10$&\multirow{2}{*}{$0.0642\pm0.0031$}&\multirow{2}{*}{$-0.579\pm0.041$}&$0.35\pm0.13$\\&$0.0636\pm0.0030$&$-0.58\pm0.04$&$0.40\pm0.07$&$-0.71\pm0.04$&$0.57\pm0.04$&$4.42\pm0.11$&$2.54\pm0.10$&&&$-0.743\pm0.071$\\
\multirow{2}{*}{$\Lambda_c^+\to\Sigma^0 K^+$}&$0.0380\pm0.0023$&$-0.64\pm0.08$&$0.77\pm0.06$&$0.01\pm0.11$&$0.83\pm0.06$&$2.91\pm0.14$&$2.26\pm0.10$&\multirow{2}{*}{$0.0370\pm0.0031$}&\multirow{2}{*}{$-0.54\pm0.20$}\\&$0.0379\pm0.0023$&$-0.61\pm0.08$&$0.79\pm0.06$&$0.03\pm0.10$&$0.84\pm0.06$&$2.88\pm0.13$&$2.22\pm0.10$\\
\multirow{2}{*}{$\Lambda_c^+\to\Sigma^+K_S$}&$0.0381\pm0.0023$&$-0.64\pm0.08$&$0.77\pm0.06$&$0.01\pm0.11$&$0.83\pm0.06$&$2.91\pm0.14$&$2.26\pm0.10$&\multirow{2}{*}{$0.047\pm0.014$}&\\&$0.0379\pm0.0023$&$-0.61\pm0.08$&$0.79\pm0.06$&$0.03\pm0.10$&$0.84\pm0.06$&$2.88\pm0.13$&$2.22\pm0.10$\\
\multirow{2}{*}{$\Lambda_c^+\to n\pi^+$}&$0.071\pm0.007$&$-0.52\pm0.11$&$-0.71\pm0.04$&$0.47\pm0.10$&$1.30\pm0.07$&$1.87\pm0.24$&$-2.21\pm0.11$&\multirow{2}{*}{$0.066\pm0.013$}&\\&$0.073\pm0.007$&$-0.56\pm0.10$&$-0.70\pm0.04$&$0.44\pm0.10$&$1.30\pm0.06$&$1.95\pm0.23$&$-2.24\pm0.10$\\
\multirow{2}{*}{$\Lambda_c^+\to p\pi^0$}&$0.0186\pm0.0034$&$-0.33\pm0.57$&$-0.94\pm0.22$&$-0.03\pm0.55$&$0.54\pm0.14$&$1.34\pm0.43$&$-1.91\pm0.61$&\multirow{2}{*}{$0.0156_{-0.0061}^{+0.0075}$}&\\&$0.0196\pm0.0061$&$-0.51\pm0.16$&$-0.76\pm0.64$&$-0.40\pm1.33$&$0.43\pm0.43$&$1.59\pm0.98$&$-2.16\pm0.35$\\
\multirow{2}{*}{$\Lambda_c^+\to p\eta$}&$0.163\pm0.010$&$-0.66\pm0.06$&$0.44\pm0.22$&$-0.61\pm0.18$&$1.08\pm0.26$&$5.63\pm0.31$&$2.55\pm0.23$&\multirow{2}{*}{$0.158\pm0.011$}&\\&$0.156\pm0.009$&$-0.66\pm0.08$&$0.24\pm0.92$&$-0.71\pm0.35$&$0.91\pm0.55$&$5.70\pm0.62$&$2.79\pm1.21$\\
\multirow{2}{*}{$\Lambda_c^+\to p\eta'$}&$0.052\pm0.008$&$-0.44\pm0.10$&$0.65\pm0.19$&$-0.61\pm0.20$&$0.73\pm0.18$&$4.82\pm0.55$&$2.17\pm0.19$&\multirow{2}{*}{$0.048\pm0.009$}&\\&$0.050\pm0.008$&$-0.46\pm0.43$&$0.61\pm0.26$&$-0.65\pm0.24$&$0.68\pm0.23$&$4.76\pm0.59$&$2.22\pm0.61$\\

\hline
Channel
&$10^{2}\mathcal{R}_X$&$\alpha$~~~~~~~&$\beta$~~~~~~~&$\gamma$~~~~~~
&$|A|$~~~~&$|B|$~~~~ &$\delta_P-\delta_S$~~~ 
&$10^{2}(\mathcal{R}_X)_\text{exp}$
&$\alpha_\text{exp}$
&$\beta_\text{exp}$
\\
\hline
\multirow{2}{*}{$\Xi_c^0\to\Lambda K_S^0$}&$23.0\pm0.8$&$-0.62\pm0.03$&$0.53\pm0.11$&$-0.58\pm0.11$&$2.47\pm0.33$&$13.57\pm0.49$&$2.43\pm0.10$&\multirow{2}{*}{$22.9\pm1.3$}&\\&$23.1\pm0.8$&$-0.61\pm0.03$&$0.47\pm0.33$&$-0.64\pm0.22$&$2.30\pm0.74$&$13.80\pm0.77$&$2.49\pm0.36$\\
\multirow{2}{*}{$\Xi_c^0\to\Sigma^0 K_S^0$}&$3.8\pm0.6$&$-0.41\pm0.60$&$-0.84\pm0.08$&$0.35\pm0.56$&$1.80\pm0.45$&$3.84\pm1.61$&$-2.02\pm0.61$&\multirow{2}{*}{$3.8\pm0.7$}&\\&$3.6\pm0.7$&$-0.63\pm0.17$&$-0.77\pm0.25$&$-0.06\pm1.64$&$1.47\pm1.37$&$4.80\pm3.47$&$-2.26\pm0.29$\\
\multirow{2}{*}{$\Xi_c^0\to\Sigma^+K^-$}&$13.9\pm1.0$&$-0.04\pm0.12$&$-0.99\pm0.02$&$-0.14\pm0.09$&$2.76\pm0.18$&$9.71\pm0.47$&$-1.61\pm0.12$&\multirow{2}{*}{$12.3\pm1.2$}&\\&$14.0\pm0.9$&$-0.06\pm0.12$&$-0.99\pm0.02$&$-0.14\pm0.09$&$2.77\pm0.18$&$9.75\pm0.46$&$-1.63\pm0.12$\\
\hline
\multirow{2}{*}{$\Xi_c^0\to\Xi^-K^+$}&$4.39\pm0.02$&$-0.72\pm0.03$&$0.66\pm0.03$&$0.21\pm0.04$&$1.86\pm0.05$&$5.44\pm0.13$&$2.40\pm0.04$&\multirow{2}{*}{$2.75\pm0.57$}&\\&$4.39\pm0.02$&$-0.71\pm0.03$&$0.67\pm0.03$&$0.22\pm0.04$&$1.86\pm0.05$&$5.41\pm0.13$&$2.38\pm0.04$\\\hline
\hline
\end{tabular}
}
\end{table*}
%%%%%%%%%

%%%%%%%%%%%%%%%%%
\begin{table*}[h]
\caption{
Same as Table \ref{tab:resultScenI} except in Case II in which the experimental data from Belle \cite{Belle-II:2024jql}  with an asterisk are included in the fit.
 }
\label{tab:resultScenII}
\centering
\resizebox{\textwidth}{!} 
{
\begin{tabular}
{ l |c rrr rrr |  c c  c }
\hline
\hline
\multirow{2}{*}{Channel}
&\multirow{2}{*}{$10^{2}\mathcal{B}$}
&\multirow{2}{*}{$\alpha$}~~~~~~~
&\multirow{2}{*}{$\beta$}~~~~~~~
&\multirow{2}{*}{$\gamma$}~~~~~~
&\multirow{2}{*}{$|A|$}~~~~
&\multirow{2}{*}{$|B|$}~~~~ 
&\multirow{2}{*}{$\delta_P-\delta_S$}~~~ 
&\multirow{2}{*}{$10^{2}\mathcal{B}_\text{exp}$}
&\multirow{2}{*}{$\alpha_\text{exp}$}
&$\beta_\text{exp}$\\
&&&&&&&&&&$\gamma_\text{exp}$\\
\hline
\multirow{2}{*}
{$\Lambda_c^+\to\Lambda^0\pi^+$}&$1.30\pm0.05$&$-0.76\pm0.01$&$0.39\pm0.02$&$0.51\pm0.01$&$5.57\pm0.10$&$9.23\pm0.20$&$2.67\pm0.02$&\multirow{2}{*}{$1.29\pm{0.05}$}&\multirow{2}{*}{$-0.762\pm0.006$}&\\&$1.30\pm0.04$&$-0.76\pm0.01$&$0.39\pm0.02$&$0.51\pm0.01$&$5.58\pm0.10$&$9.26\pm0.20$&$2.67\pm0.02$\\
\multirow{2}{*}{$\Lambda_c^+\to\Sigma^0\pi^+$}&$1.24\pm0.05$&$-0.47\pm0.01$&$0.32\pm0.10$&$-0.82\pm0.04$&$1.86\pm0.20$&$19.13\pm0.44$&$2.54\pm0.14$&\multirow{2}{*}{$1.27\pm{0.06}$}&\multirow{2}{*}{$-0.466\pm0.018$}\\&$1.24\pm0.05$&$-0.47\pm0.01$&$0.33\pm0.10$&$-0.82\pm0.04$&$1.88\pm0.20$&$19.12\pm0.44$&$2.53\pm0.14$\\
\multirow{2}{*}{$\Lambda_c^+\to\Sigma^+ \pi^0$}&$1.25\pm0.05$&$-0.47\pm0.01$&$0.32\pm0.10$&$-0.82\pm0.04$&$1.86\pm0.20$&$19.13\pm0.44$&$2.54\pm0.14$&\multirow{2}{*}{$1.24\pm0.09$}&\multirow{2}{*}{$-0.484\pm0.027$}\\&$1.25\pm0.05$&$-0.47\pm0.01$&$0.33\pm0.10$&$-0.82\pm0.04$&$1.88\pm0.20$&$19.12\pm0.44$&$2.53\pm0.14$\\
\multirow{2}{*}{$\Lambda_c^+\to\Sigma^+ \eta$}&$0.33\pm0.04$&$-0.90\pm0.04$&$-0.14\pm0.12$&$0.42\pm0.10$&$2.96\pm0.19$&$6.83\pm0.69$&$-2.99\pm0.13$&\multirow{2}{*}{$0.32\pm0.05$}&\multirow{2}{*}{$-0.99\pm0.06$}\\&$0.33\pm0.04$&$-0.89\pm0.04$&$-0.08\pm0.12$&$0.44\pm0.09$&$2.97\pm0.19$&$6.68\pm0.69$&$-3.06\pm0.13$\\
\multirow{2}{*}{$\Lambda_c^+\to\Sigma^+ \eta'$}&$0.18\pm0.03$&$-0.44\pm0.07$&$0.84\pm0.13$&$-0.32\pm0.34$&$2.12\pm0.67$&$18.43\pm1.92$&$2.05\pm0.10$&\multirow{2}{*}{$0.41\pm0.08$}&\multirow{2}{*}{$-0.46\pm0.07$}\\&$0.17\pm0.04$&$-0.44\pm0.07$&$0.88\pm0.08$&$-0.16\pm0.38$&$2.29\pm0.67$&$16.85\pm2.41$&$2.03\pm0.08$\\
\multirow{2}{*}{$\Lambda_c^+\to\Xi^0 K^+$}&$0.33\pm0.03$&$-0.08\pm0.12$&$-0.98\pm0.02$&$0.18\pm0.09$&$2.72\pm0.18$&$9.69\pm0.47$&$-1.65\pm0.12$&\multirow{2}{*}{$0.55\pm0.07$}&\multirow{2}{*}{$0.01\pm0.16$}\\&$0.33\pm0.03$&$-0.07\pm0.11$&$-0.98\pm0.02$&$0.18\pm0.09$&$2.72\pm0.18$&$9.69\pm0.46$&$-1.64\pm0.11$\\
\multirow{2}{*}{$\Lambda_c^+\to p K_S$}&$1.59\pm0.06$&$-0.74\pm0.03$&$0.21\pm0.59$&$-0.63\pm0.20$&$3.22\pm0.91$&$17.31\pm1.02$&$2.86\pm0.74$&\multirow{2}{*}{$1.59\pm0.07$}&\multirow{2}{*}{$-0.743\pm0.028$}\\&$1.56\pm0.06$&$-0.74\pm0.03$&$0.63\pm0.05$&$-0.24\pm0.10$&$4.60\pm0.31$&$14.92\pm0.67$&$2.44\pm0.05$\\
\multirow{2}{*}{$\Xi_c^0\to\Xi^- \pi^+$}&$2.97\pm0.08$&$-0.74\pm0.03$&$0.66\pm0.03$&$0.11\pm0.04$&$8.02\pm0.20$&$23.76\pm0.54$&$2.42\pm0.04$&\multirow{2}{*}{$1.80\pm0.52$}&\multirow{2}{*}{$-0.64\pm0.05$}\\&$2.99\pm0.08$&$-0.75\pm0.03$&$0.66\pm0.03$&$0.12\pm0.04$&$8.06\pm0.20$&$23.80\pm0.52$&$2.42\pm0.04$\\
\multirow{2}{*}{$\Xi_c^0\to\Xi^0 \pi^0$}&$0.71\pm0.04$&$-0.69\pm0.07$&$0.72\pm0.07$&$-0.09\pm0.10$&$3.53\pm0.24$&$12.80\pm0.57$&$2.34\pm0.10$&\multirow{2}{*}{$0.69\pm0.16^{*}$}&\multirow{2}{*}{$-0.90\pm0.27^{*}$}\\&$0.72\pm0.04$&$-0.69\pm0.06$&$0.71\pm0.06$&$-0.09\pm0.10$&$3.57\pm0.23$&$12.81\pm0.56$&$2.34\pm0.09$\\
\multirow{2}{*}{$\Xi_c^0\to\Xi^0 \eta$}&$0.25\pm0.03$&$-0.03\pm0.14$&$-0.19\pm0.13$&$-0.98\pm0.03$&$0.33\pm0.22$&$12.38\pm0.76$&$-1.73\pm0.72$&\multirow{2}{*}{$0.16\pm0.05^{*}$}\\&$0.26\pm0.03$&$-0.01\pm0.13$&$-0.17\pm0.13$&$-0.99\pm0.02$&$0.29\pm0.22$&$12.62\pm0.62$&$-1.63\pm0.74$\\
\multirow{2}{*}{$\Xi_c^0\to\Xi^0 \eta'$}&$0.23\pm0.03$&$-0.71\pm0.07$&$0.46\pm0.22$&$-0.53\pm0.23$&$2.03\pm0.54$&$20.86\pm1.91$&$2.56\pm0.21$&\multirow{2}{*}{$0.12\pm0.04^{*}$}\\&$0.21\pm0.03$&$-0.74\pm0.07$&$0.53\pm0.19$&$-0.42\pm0.29$&$2.18\pm0.59$&$19.25\pm2.40$&$2.52\pm0.16$\\
\multirow{2}{*}{$\Xi_c^+\to\Xi^0\pi^+$}&$0.96\pm0.11$&$-0.80\pm0.09$&$0.38\pm0.11$&$0.46\pm0.11$&$3.02\pm0.22$&$6.05\pm0.71$&$2.70\pm0.15$&\multirow{2}{*}{$1.6\pm0.8$}&\\&$0.95\pm0.11$&$-0.79\pm0.08$&$0.39\pm0.10$&$0.46\pm0.10$&$3.01\pm0.21$&$6.03\pm0.63$&$2.68\pm0.13$\\
\hline
\multirow{2}{*}{$\Lambda_c^+\to\Lambda^0 K^+$}&$0.0628\pm0.0030$&$-0.57\pm0.04$&$0.42\pm0.07$&$-0.70\pm0.04$&$0.58\pm0.04$&$4.37\pm0.11$&$2.51\pm0.10$&\multirow{2}{*}{$0.0642\pm0.0031$}&\multirow{2}{*}{$-0.579\pm0.041$}&\\&$0.0629\pm0.0030$&$-0.57\pm0.04$&$0.43\pm0.07$&$-0.70\pm0.04$&$0.58\pm0.04$&$4.38\pm0.11$&$2.50\pm0.09$\\
\multirow{2}{*}{$\Lambda_c^+\to\Sigma^0 K^+$}&$0.0376\pm0.0022$&$-0.69\pm0.07$&$0.72\pm0.07$&$-0.03\pm0.10$&$0.81\pm0.05$&$2.95\pm0.13$&$2.34\pm0.10$&\multirow{2}{*}{$0.0370\pm0.0031$}&\multirow{2}{*}{$-0.54\pm0.20$}\\&$0.0380\pm0.0022$&$-0.70\pm0.06$&$0.72\pm0.06$&$-0.02\pm0.10$&$0.82\pm0.05$&$2.95\pm0.13$&$2.34\pm0.09$\\
\multirow{2}{*}{$\Lambda_c^+\to\Sigma^+K_S$}&$0.0376\pm0.0022$&$-0.69\pm0.07$&$0.72\pm0.07$&$-0.03\pm0.10$&$0.81\pm0.05$&$2.95\pm0.13$&$2.34\pm0.10$&\multirow{2}{*}{$0.047\pm0.014$}&\\&$0.0380\pm0.0021$&$-0.70\pm0.06$&$0.72\pm0.06$&$-0.02\pm0.10$&$0.82\pm0.05$&$2.95\pm0.13$&$2.34\pm0.09$\\
\multirow{2}{*}{$\Lambda_c^+\to n\pi^+$}&$0.069\pm0.007$&$-0.47\pm0.11$&$-0.70\pm0.03$&$0.53\pm0.10$&$1.31\pm0.06$&$1.75\pm0.24$&$-2.16\pm0.12$&\multirow{2}{*}{$0.066\pm0.013$}&\\&$0.069\pm0.007$&$-0.47\pm0.11$&$-0.70\pm0.03$&$0.53\pm0.10$&$1.31\pm0.06$&$1.73\pm0.23$&$-2.16\pm0.11$\\
\multirow{2}{*}{$\Lambda_c^+\to p\pi^0$}&$0.0212\pm0.0046$&$-0.30\pm0.94$&$-0.65\pm0.19$&$-0.69\pm0.56$&$0.32\pm0.27$&$1.82\pm0.46$&$-2.01\pm1.10$&\multirow{2}{*}{$0.0156_{-0.0061}^{+0.0075}$}&\\&$0.0171\pm0.0030$&$0.77\pm0.30$&$-0.53\pm0.37$&$0.37\pm0.15$&$0.61\pm0.05$&$1.00\pm0.17$&$-0.60\pm0.50$\\
\multirow{2}{*}{$\Lambda_c^+\to p\eta$}&$0.156\pm0.010$&$-0.59\pm0.07$&$0.03\pm0.72$&$-0.81\pm0.04$&$0.74\pm0.09$&$5.86\pm0.19$&$3.09\pm1.22$&\multirow{2}{*}{$0.158\pm0.011$}&\\&$0.167\pm0.007$&$-0.66\pm0.04$&$0.64\pm0.05$&$-0.39\pm0.08$&$1.37\pm0.09$&$5.32\pm0.19$&$2.37\pm0.06$\\
\multirow{2}{*}{$\Lambda_c^+\to p\eta'$}&$0.037\pm0.007$&$-0.31\pm0.29$&$0.30\pm0.24$&$-0.90\pm0.13$&$0.30\pm0.20$&$4.39\pm0.49$&$2.38\pm0.59$&\multirow{2}{*}{$0.048\pm0.009$}&\\&$0.048\pm0.010$&$-0.49\pm0.13$&$0.59\pm0.21$&$-0.64\pm0.13$&$0.67\pm0.14$&$4.69\pm0.48$&$2.27\pm0.29$\\

\hline
Channel
&$10^{2}\mathcal{R}_X$&$\alpha$~~~~~~~&$\beta$~~~~~~~&$\gamma$~~~~~~
&$|A|$~~~~&$|B|$~~~~ &$\delta_P-\delta_S$~~~ 
&$10^{2}(\mathcal{R}_X)_\text{exp}$
&$\alpha_\text{exp}$
&$\beta_\text{exp}$
\\
\hline
\multirow{2}{*}{$\Xi_c^0\to\Lambda K_S^0$}&$22.8\pm0.9$&$-0.63\pm0.07$&$0.32\pm0.34$&$-0.71\pm0.10$&$2.06\pm0.37$&$13.99\pm0.34$&$2.67\pm0.48$&\multirow{2}{*}{$22.9\pm1.3$}&\\&$23.0\pm0.8$&$-0.67\pm0.02$&$0.53\pm0.07$&$-0.53\pm0.07$&$2.64\pm0.19$&$13.34\pm0.37$&$2.47\pm0.07$\\
\multirow{2}{*}{$\Xi_c^0\to\Sigma^0 K_S^0$}&$3.6\pm0.6$&$-0.45\pm1.02$&$-0.77\pm0.17$&$-0.46\pm0.82$&$1.11\pm0.85$&$5.61\pm1.56$&$-2.10\pm1.06$&\multirow{2}{*}{$3.8\pm0.7$}&\\&$3.8\pm0.6$&$0.58\pm0.28$&$-0.37\pm0.31$&$0.72\pm0.10$&$2.05\pm0.17$&$2.53\pm0.55$&$-0.57\pm0.59$\\
\multirow{2}{*}{$\Xi_c^0\to\Sigma^+K^-$}&$13.7\pm0.9$&$-0.08\pm0.11$&$-0.99\pm0.02$&$-0.15\pm0.09$&$2.72\pm0.18$&$9.69\pm0.47$&$-1.65\pm0.12$&\multirow{2}{*}{$12.3\pm1.2$}&\\&$13.6\pm0.9$&$-0.07\pm0.11$&$-0.99\pm0.02$&$-0.15\pm0.09$&$2.72\pm0.18$&$9.69\pm0.46$&$-1.64\pm0.11$\\
\hline
\multirow{2}{*}{$\Xi_c^0\to\Xi^-K^+$}&$4.38\pm0.01$&$-0.73\pm0.03$&$0.65\pm0.03$&$0.20\pm0.04$&$1.85\pm0.05$&$5.47\pm0.12$&$2.42\pm0.04$&\multirow{2}{*}{$2.75\pm0.57$}&\\&$4.38\pm0.01$&$-0.74\pm0.03$&$0.65\pm0.03$&$0.20\pm0.04$&$1.86\pm0.05$&$5.48\pm0.12$&$2.42\pm0.04$\\
\hline
\hline
\end{tabular}
}
\end{table*}
%%%%%%%%%%%%%%%%%

%%%%%%%%%%%%%%%%%%%
\begin{table*}[tp!]\footnotesize
\caption{
Same as Table \ref{tab:resultScenI} except for yet-observed CF and SCS  modes in Case I.
}
\label{tab:fitotherScenI}
 \resizebox{\textwidth}{!} 
 {
\centering
\begin{tabular}
{ l | r rrr rr r
}
\hline
\hline
Channel&
$10^{3}\mathcal{B}$~~~~ &$\alpha$~~~~~~~&$\beta$~~~~~~~&$\gamma$~~~~~~
&$|A|$~~~ & $|B|$~~~ & $\delta_P-\delta_S$~~~\\
\hline
% CF
\multirow{2}{*}
{$\Lambda_c^{+} \rightarrow p K_L$}&$15.02\pm0.56$&$-0.73\pm0.03$&$0.57\pm0.14$&$-0.37\pm0.20$&$4.11\pm0.64$&$15.38\pm1.22$&$2.48\pm0.13$\\&$15.22\pm0.57$&$-0.74\pm0.04$&$0.47\pm0.51$&$-0.49\pm0.45$&$3.72\pm1.65$&$16.15\pm2.35$&$2.58\pm0.51$\\
\multirow{2}{*}{$\Xi_c^0 \rightarrow \Lambda^0 K_L$}&$7.64\pm0.21$&$-0.59\pm0.03$&$0.50\pm0.10$&$-0.63\pm0.09$&$2.45\pm0.32$&$14.56\pm0.46$&$2.43\pm0.10$\\&$7.64\pm0.28$&$-0.58\pm0.03$&$0.45\pm0.29$&$-0.67\pm0.18$&$2.31\pm0.67$&$14.74\pm0.67$&$2.48\pm0.33$\\
\multirow{2}{*}{$\Xi_c^0 \rightarrow \Sigma^0 K_L$}&$1.01\pm0.16$&$-0.24\pm0.59$&$-0.97\pm0.13$&$0.04\pm0.48$&$1.50\pm0.40$&$4.44\pm1.07$&$-1.82\pm0.60$\\&$0.96\pm0.16$&$-0.45\pm0.17$&$-0.84\pm0.55$&$-0.31\pm1.39$&$1.20\pm1.25$&$5.07\pm2.57$&$-2.07\pm0.39$\\
\multirow{2}{*}{$\Xi_c^{+} \rightarrow \Sigma^{+} K_S$}&$1.77\pm0.75$&$-0.87\pm0.66$&$-0.16\pm0.49$&$-0.47\pm1.10$&$0.82\pm0.73$&$4.21\pm2.32$&$-2.95\pm0.66$\\&$2.80\pm4.37$&$-0.30\pm2.11$&$-0.44\pm0.84$&$-0.85\pm0.38$&$0.56\pm0.33$&$5.94\pm5.19$&$-2.18\pm4.12$\\
\hline
\multirow{2}{*}{$\Xi_c^0 \rightarrow \Sigma^{+} \pi^{-}$}&$0.26\pm0.02$&$-0.04\pm0.12$&$-0.98\pm0.02$&$-0.20\pm0.09$&$0.64\pm0.04$&$2.24\pm0.11$&$-1.61\pm0.12$\\&$0.26\pm0.02$&$-0.06\pm0.12$&$-0.98\pm0.02$&$-0.20\pm0.09$&$0.64\pm0.04$&$2.25\pm0.11$&$-1.63\pm0.12$\\
\multirow{2}{*}{$\Xi_c^0 \rightarrow \Sigma^0 \pi^0 $}&$0.33\pm0.03$&$-0.78\pm0.19$&$0.39\pm0.10$&$0.49\pm0.24$&$0.97\pm0.11$&$1.64\pm0.36$&$2.68\pm0.20$\\&$0.31\pm0.08$&$-0.87\pm0.34$&$0.35\pm0.28$&$0.33\pm0.60$&$0.90\pm0.31$&$1.83\pm0.61$&$2.76\pm0.41$\\
\multirow{2}{*}{$\Xi_c^0 \rightarrow \Sigma^0 \eta$}&$0.16\pm0.03$&$-1.00\pm0.06$&$0.04\pm0.17$&$-0.08\pm0.64$&$0.57\pm0.17$&$1.95\pm0.73$&$3.10\pm0.18$\\&$0.18\pm0.08$&$-0.86\pm0.80$&$-0.06\pm0.28$&$-0.51\pm1.30$&$0.44\pm0.50$&$2.44\pm1.59$&$-3.07\pm0.39$\\
\multirow{2}{*}{$\Xi_c^0 \rightarrow \Sigma^0 \eta'$}&$0.18\pm0.03$&$-0.49\pm0.11$&$0.86\pm0.06$&$-0.16\pm0.25$&$0.70\pm0.13$&$3.45\pm0.47$&$2.08\pm0.12$\\&$0.21\pm0.03$&$-0.52\pm0.21$&$0.79\pm0.18$&$-0.33\pm0.26$&$0.67\pm0.15$&$3.92\pm0.49$&$2.15\pm0.28$\\
\multirow{2}{*}{$\Xi_c^0 \rightarrow \Sigma^- \pi^{+}$}&$1.82\pm0.05$&$-0.73\pm0.03$&$0.68\pm0.03$&$-0.02\pm0.04$&$1.86\pm0.05$&$5.44\pm0.13$&$2.40\pm0.04$\\&$1.81\pm0.05$&$-0.73\pm0.03$&$0.69\pm0.03$&$-0.01\pm0.04$&$1.86\pm0.05$&$5.41\pm0.13$&$2.38\pm0.04$\\
\multirow{2}{*}{$\Xi_c^0 \rightarrow \Xi^0 K_{S / L}$}&$0.39\pm0.01$&$-0.52\pm0.02$&$0.38\pm0.11$&$-0.77\pm0.05$&$0.45\pm0.05$&$4.42\pm0.11$&$2.50\pm0.14$\\&$0.39\pm0.01$&$-0.51\pm0.02$&$0.40\pm0.11$&$-0.76\pm0.05$&$0.45\pm0.05$&$4.41\pm0.11$&$2.48\pm0.13$\\
\multirow{2}{*}{$\Xi_c^0 \rightarrow p K^{-}$}&$0.31\pm0.02$&$-0.04\pm0.11$&$-0.92\pm0.04$&$-0.39\pm0.08$&$0.64\pm0.04$&$2.24\pm0.11$&$-1.61\pm0.12$\\&$0.31\pm0.02$&$-0.05\pm0.11$&$-0.92\pm0.04$&$-0.39\pm0.08$&$0.64\pm0.04$&$2.25\pm0.11$&$-1.63\pm0.12$\\
\multirow{2}{*}{$\Xi_c^0 \rightarrow n K_{S / L}$}&$0.88\pm0.04$&$-0.36\pm0.01$&$0.27\pm0.08$&$-0.90\pm0.03$&$0.45\pm0.05$&$4.42\pm0.11$&$2.50\pm0.14$\\&$0.87\pm0.04$&$-0.36\pm0.01$&$0.28\pm0.08$&$-0.89\pm0.03$&$0.45\pm0.05$&$4.41\pm0.11$&$2.48\pm0.13$\\
\multirow{2}{*}{$\Xi_c^0 \rightarrow \Lambda^0 \pi^0 $}&$0.06\pm0.01$&$-0.26\pm0.44$&$0.40\pm0.17$&$0.88\pm0.13$&$0.48\pm0.07$&$0.32\pm0.16$&$2.15\pm0.84$\\&$0.07\pm0.05$&$0.06\pm0.88$&$0.40\pm0.17$&$0.92\pm0.11$&$0.52\pm0.17$&$0.29\pm0.27$&$1.42\pm2.15$\\
\multirow{2}{*}{$\Xi_c^0 \rightarrow \Lambda^0 \eta$}&$0.39\pm0.06$&$-0.44\pm0.25$&$0.83\pm0.13$&$0.35\pm0.13$&$1.07\pm0.11$&$2.12\pm0.23$&$2.06\pm0.29$\\&$0.40\pm0.14$&$-0.52\pm0.38$&$0.84\pm0.19$&$0.18\pm0.29$&$1.01\pm0.29$&$2.42\pm0.26$&$2.13\pm0.43$\\
\multirow{2}{*}{$\Xi_c^0 \rightarrow \Lambda^0 \eta'$}&$0.64\pm0.08$&$-0.61\pm0.08$&$0.62\pm0.15$&$-0.50\pm0.22$&$0.99\pm0.21$&$6.14\pm0.67$&$2.35\pm0.12$\\&$0.76\pm0.15$&$-0.54\pm0.11$&$0.55\pm0.16$&$-0.64\pm0.20$&$0.92\pm0.21$&$7.03\pm1.01$&$2.35\pm0.15$\\
\multirow{2}{*}{$\Xi_c^{+} \rightarrow p K_{S / L}$}&$1.55\pm0.08$&$-0.59\pm0.06$&$0.71\pm0.08$&$-0.38\pm0.09$&$0.83\pm0.06$&$2.91\pm0.14$&$2.26\pm0.10$\\&$1.53\pm0.08$&$-0.56\pm0.07$&$0.74\pm0.07$&$-0.37\pm0.09$&$0.84\pm0.06$&$2.88\pm0.13$&$2.22\pm0.10$\\
\multirow{2}{*}{$\Xi_c^{+} \rightarrow \Sigma^{+} \pi^0$}&$2.62\pm0.12$&$-0.54\pm0.11$&$0.26\pm0.16$&$-0.80\pm0.12$&$0.59\pm0.16$&$5.03\pm0.21$&$2.69\pm0.20$\\&$2.70\pm0.21$&$-0.51\pm0.13$&$0.14\pm0.56$&$-0.85\pm0.17$&$0.51\pm0.27$&$5.18\pm0.42$&$2.87\pm0.96$\\
\multirow{2}{*}{$\Xi_c^{+} \rightarrow \Sigma^{+} \eta$}&$0.97\pm0.21$&$-1.00\pm0.06$&$0.04\pm0.17$&$-0.09\pm0.64$&$0.81\pm0.24$&$2.76\pm1.03$&$3.10\pm1.15$\\&$1.09\pm0.51$&$-0.86\pm0.80$&$-0.06\pm0.28$&$-0.51\pm1.30$&$0.63\pm0.70$&$3.46\pm2.25$&$-3.07\pm0.39$\\
\multirow{2}{*}{$\Xi_c^{+} \rightarrow \Sigma^{+} \eta'$}&$1.08\pm0.19$&$-0.48\pm0.11$&$0.86\pm0.06$&$-0.16\pm0.25$&$0.99\pm0.18$&$4.88\pm0.66$&$2.08\pm0.12$\\&$1.23\pm0.21$&$-0.52\pm0.21$&$0.79\pm0.18$&$-0.33\pm0.26$&$0.94\pm0.21$&$5.55\pm0.69$&$2.15\pm0.28$\\
\multirow{2}{*}{$\Xi_c^{+} \rightarrow \Sigma^0 \pi^{+}$}&$3.20\pm0.10$&$-0.59\pm0.02$&$0.53\pm0.05$&$-0.61\pm0.04$&$0.90\pm0.05$&$5.28\pm0.10$&$2.41\pm0.06$\\&$3.19\pm0.10$&$-0.59\pm0.02$&$0.54\pm0.05$&$-0.60\pm0.04$&$0.91\pm0.05$&$5.26\pm0.10$&$2.41\pm0.06$\\
\multirow{2}{*}{$\Xi_c^{+} \rightarrow \Xi^0 K^{+}$}&$1.34\pm0.13$&$-0.41\pm0.10$&$-0.55\pm0.03$&$0.72\pm0.06$&$1.30\pm0.07$&$1.87\pm0.24$&$-2.21\pm0.11$\\&$1.34\pm0.12$&$-0.44\pm0.09$&$-0.56\pm0.03$&$0.70\pm0.06$&$1.30\pm0.06$&$1.95\pm0.23$&$-2.24\pm0.10$\\
\multirow{2}{*}{$\Xi_c^{+} \rightarrow \Lambda^0 \pi^{+}$}&$0.17\pm0.04$&$-0.20\pm0.45$&$0.45\pm0.31$&$0.87\pm0.13$&$0.45\pm0.06$&$0.31\pm0.14$&$1.99\pm0.57$\\&$0.17\pm0.04$&$-0.05\pm0.45$&$0.47\pm0.29$&$0.88\pm0.15$&$0.46\pm0.06$&$0.31\pm0.18$&$1.68\pm0.99$\\
\hline
\hline
\end{tabular}
 }
\end{table*}
%%%%%%%%%%%

%%%%%%%%%%%%%%%%%%
\begin{table*}[tp!]\footnotesize
\caption{
Same as Table \ref{tab:fitotherScenI} except for yet-observed CF and SCS  modes in Case II.
}
\label{tab:fitotherScenII} 
\resizebox{\textwidth}{!} 
 {
\centering
\begin{tabular}
{ l | r rrr rr r
}
\hline
\hline
Channel&
$10^{3}\mathcal{B}$~~~~ &$\alpha$~~~~~~~&$\beta$~~~~~~~&$\gamma$~~~~~~
&$|A|$~~~ & $|B|$~~~ & $\delta_P-\delta_S$~~~\\
\hline
% CF
\multirow{2}{*}
{$\Lambda_c^{+} \rightarrow p K_L$}&$15.23\pm0.64$&$-0.75\pm0.03$&$0.26\pm0.54$&$-0.61\pm0.20$&$3.24\pm0.87$&$16.82\pm0.90$&$2.81\pm0.66$\\&$15.00\pm0.59$&$-0.74\pm0.03$&$0.63\pm0.05$&$-0.25\pm0.09$&$4.48\pm0.29$&$14.68\pm0.63$&$2.44\pm0.05$\\
\multirow{2}{*}{$\Xi_c^0 \rightarrow \Lambda^0 K_L$}&$7.55\pm0.25$&$-0.60\pm0.07$&$0.32\pm0.30$&$-0.73\pm0.08$&$2.09\pm0.34$&$14.90\pm0.33$&$2.65\pm0.43$\\&$7.65\pm0.21$&$-0.64\pm0.02$&$0.50\pm0.07$&$-0.59\pm0.06$&$2.60\pm0.19$&$14.35\pm0.37$&$2.48\pm0.07$\\
\multirow{2}{*}{$\Xi_c^0 \rightarrow \Sigma^0 K_L$}&$0.96\pm0.15$&$-0.26\pm1.02$&$-0.73\pm0.25$&$-0.63\pm0.69$&$0.87\pm0.82$&$5.66\pm1.27$&$-1.92\pm1.15$\\&$1.00\pm0.16$&$0.75\pm0.27$&$-0.55\pm0.32$&$0.36\pm0.12$&$1.71\pm0.15$&$3.63\pm0.48$&$-0.63\pm0.45$\\
\multirow{2}{*}{$\Xi_c^{+} \rightarrow \Sigma^{+} K_S$}&$3.58\pm4.07$&$0.18\pm1.07$&$-0.67\pm0.82$&$-0.72\pm1.00$&$0.84\pm2.00$&$6.49\pm1.94$&$-1.31\pm1.22$\\&$3.23\pm1.16$&$-0.70\pm0.20$&$0.65\pm0.22$&$0.28\pm0.25$&$1.73\pm0.32$&$3.98\pm1.07$&$2.39\pm0.30$\\
\hline
\multirow{2}{*}{$\Xi_c^0 \rightarrow \Sigma^{+} \pi^{-}$}&$0.26\pm0.02$&$-0.08\pm0.11$&$-0.97\pm0.02$&$-0.21\pm0.09$&$0.63\pm0.04$&$2.23\pm0.11$&$-1.65\pm0.12$\\&$0.26\pm0.02$&$-0.07\pm0.11$&$-0.97\pm0.02$&$-0.21\pm0.09$&$0.63\pm0.04$&$2.23\pm0.11$&$-1.64\pm0.11$\\
\multirow{2}{*}{$\Xi_c^0 \rightarrow \Sigma^0 \pi^0 $}&$0.30\pm0.06$&$-0.98\pm0.11$&$0.16\pm0.44$&$0.10\pm0.38$&$0.80\pm0.21$&$2.10\pm0.28$&$2.98\pm0.45$\\&$0.32\pm0.03$&$-0.61\pm0.08$&$0.47\pm0.07$&$0.63\pm0.08$&$1.01\pm0.05$&$1.38\pm0.14$&$2.48\pm0.11$\\
\multirow{2}{*}{$\Xi_c^0 \rightarrow \Sigma^0 \eta$}&$0.19\pm0.04$&$-0.66\pm0.84$&$-0.30\pm0.79$&$-0.69\pm0.48$&$0.36\pm0.26$&$2.63\pm0.58$&$-2.72\pm1.46$\\&$0.13\pm0.02$&$-0.67\pm0.20$&$0.18\pm0.22$&$0.72\pm0.16$&$0.71\pm0.07$&$0.89\pm0.28$&$2.88\pm0.35$\\
\multirow{2}{*}{$\Xi_c^0 \rightarrow \Sigma^0 \eta'$}&$0.09\pm0.01$&$-0.59\pm0.32$&$0.61\pm0.24$&$-0.53\pm0.28$&$0.38\pm0.11$&$2.83\pm0.36$&$2.33\pm0.42$\\&$0.12\pm0.02$&$-0.45\pm0.07$&$0.81\pm0.11$&$-0.37\pm0.27$&$0.48\pm0.11$&$2.99\pm0.35$&$2.08\pm0.08$\\
\multirow{2}{*}{$\Xi_c^0 \rightarrow \Sigma^- \pi^{+}$}&$1.82\pm0.05$&$-0.75\pm0.03$&$0.66\pm0.03$&$-0.03\pm0.04$&$1.85\pm0.05$&$5.47\pm0.12$&$2.42\pm0.04$\\&$1.83\pm0.05$&$-0.75\pm0.03$&$0.66\pm0.03$&$-0.03\pm0.04$&$1.86\pm0.05$&$5.48\pm0.12$&$2.42\pm0.04$\\
\multirow{2}{*}{$\Xi_c^0 \rightarrow \Xi^0 K_{S / L}$}&$0.38\pm0.01$&$-0.52\pm0.02$&$0.35\pm0.11$&$-0.78\pm0.05$&$0.43\pm0.05$&$4.41\pm0.10$&$2.54\pm0.14$\\&$0.38\pm0.01$&$-0.52\pm0.02$&$0.36\pm0.11$&$-0.78\pm0.05$&$0.43\pm0.05$&$4.40\pm0.10$&$2.53\pm0.14$\\
\multirow{2}{*}{$\Xi_c^0 \rightarrow p K^{-}$}&$0.30\pm0.02$&$-0.07\pm0.11$&$-0.91\pm0.04$&$-0.40\pm0.08$&$0.63\pm0.04$&$2.23\pm0.11$&$-1.65\pm0.12$\\&$0.30\pm0.02$&$-0.07\pm0.10$&$-0.91\pm0.04$&$-0.40\pm0.08$&$0.63\pm0.04$&$2.23\pm0.11$&$-1.64\pm0.11$\\
\multirow{2}{*}{$\Xi_c^0 \rightarrow n K_{S / L}$}&$0.87\pm0.03$&$-0.36\pm0.01$&$0.24\pm0.08$&$-0.90\pm0.02$&$0.43\pm0.05$&$4.41\pm0.10$&$2.54\pm0.14$\\&$0.87\pm0.03$&$-0.36\pm0.01$&$0.25\pm0.08$&$-0.90\pm0.02$&$0.43\pm0.05$&$4.40\pm0.10$&$2.53\pm0.14$\\
\multirow{2}{*}{$\Xi_c^0 \rightarrow \Lambda^0 \pi^0 $}&$0.08\pm0.05$&$-0.06\pm0.42$&$0.39\pm0.16$&$0.92\pm0.07$&$0.56\pm0.16$&$0.31\pm0.13$&$1.71\pm1.05$\\&$0.08\pm0.02$&$-0.84\pm0.14$&$0.28\pm0.15$&$0.47\pm0.22$&$0.47\pm0.05$&$0.76\pm0.23$&$2.82\pm0.19$\\
\multirow{2}{*}{$\Xi_c^0 \rightarrow \Lambda^0 \eta$}&$0.28\pm0.07$&$-0.76\pm0.35$&$0.63\pm0.38$&$0.13\pm0.21$&$0.83\pm0.17$&$2.09\pm0.20$&$2.45\pm0.52$\\&$0.37\pm0.03$&$-0.37\pm0.08$&$0.86\pm0.06$&$0.35\pm0.12$&$1.04\pm0.07$&$2.07\pm0.21$&$1.97\pm0.10$\\
\multirow{2}{*}{$\Xi_c^0 \rightarrow \Lambda^0 \eta'$}&$0.41\pm0.06$&$-0.54\pm0.11$&$0.26\pm0.23$&$-0.80\pm0.11$&$0.50\pm0.14$&$5.41\pm0.51$&$2.69\pm0.34$\\&$0.32\pm0.07$&$-0.63\pm0.09$&$0.28\pm0.25$&$-0.73\pm0.15$&$0.52\pm0.12$&$4.72\pm0.66$&$2.73\pm0.31$\\
\multirow{2}{*}{$\Xi_c^{+} \rightarrow p K_{S / L}$}&$1.55\pm0.08$&$-0.63\pm0.06$&$0.66\pm0.08$&$-0.41\pm0.08$&$0.81\pm0.05$&$2.95\pm0.13$&$2.34\pm0.10$\\&$1.56\pm0.08$&$-0.64\pm0.05$&$0.65\pm0.07$&$-0.41\pm0.08$&$0.82\pm0.05$&$2.95\pm0.13$&$2.34\pm0.09$\\
\multirow{2}{*}{$\Xi_c^{+} \rightarrow \Sigma^{+} \pi^0$}&$2.68\pm0.18$&$-0.47\pm0.05$&$-0.03\pm0.54$&$-0.88\pm0.04$&$0.45\pm0.09$&$5.21\pm0.14$&$-3.07\pm1.14$\\&$2.70\pm0.10$&$-0.66\pm0.05$&$0.30\pm0.09$&$-0.69\pm0.06$&$0.74\pm0.07$&$4.94\pm0.14$&$2.72\pm0.12$\\
\multirow{2}{*}{$\Xi_c^{+} \rightarrow \Sigma^{+} \eta$}&$1.13\pm0.22$&$-0.66\pm0.84$&$-0.30\pm0.78$&$-0.69\pm0.48$&$0.51\pm0.36$&$3.72\pm0.82$&$-2.72\pm1.46$\\&$0.78\pm0.14$&$-0.67\pm0.20$&$0.18\pm0.22$&$0.72\pm0.16$&$1.00\pm0.10$&$1.26\pm0.40$&$2.88\pm0.35$\\
\multirow{2}{*}{$\Xi_c^{+} \rightarrow \Sigma^{+} \eta'$}&$0.56\pm0.09$&$-0.59\pm0.32$&$0.61\pm0.24$&$-0.53\pm0.28$&$0.53\pm0.16$&$4.00\pm0.51$&$2.33\pm0.42$\\&$0.69\pm0.10$&$-0.45\pm0.07$&$0.81\pm0.11$&$-0.37\pm0.27$&$0.68\pm0.16$&$4.23\pm0.50$&$2.08\pm0.08$\\
\multirow{2}{*}{$\Xi_c^{+} \rightarrow \Sigma^0 \pi^{+}$}&$3.18\pm0.09$&$-0.59\pm0.02$&$0.53\pm0.05$&$-0.61\pm0.03$&$0.90\pm0.04$&$5.26\pm0.09$&$2.42\pm0.06$\\&$3.18\pm0.09$&$-0.59\pm0.02$&$0.53\pm0.05$&$-0.60\pm0.03$&$0.91\pm0.04$&$5.26\pm0.09$&$2.41\pm0.06$\\
\multirow{2}{*}{$\Xi_c^{+} \rightarrow \Xi^0 K^{+}$}&$1.32\pm0.12$&$-0.36\pm0.10$&$-0.54\pm0.04$&$0.76\pm0.06$&$1.31\pm0.06$&$1.75\pm0.24$&$-2.16\pm0.12$\\&$1.32\pm0.12$&$-0.36\pm0.09$&$-0.54\pm0.04$&$0.76\pm0.06$&$1.31\pm0.06$&$1.73\pm0.23$&$-2.16\pm0.11$\\
\multirow{2}{*}{$\Xi_c^{+} \rightarrow \Lambda^0 \pi^{+}$}&$0.16\pm0.04$&$-0.57\pm0.37$&$0.31\pm0.30$&$0.76\pm0.22$&$0.43\pm0.06$&$0.42\pm0.21$&$2.64\pm0.57$\\&$0.17\pm0.04$&$-0.59\pm0.31$&$0.29\pm0.27$&$0.76\pm0.20$&$0.44\pm0.06$&$0.43\pm0.19$&$2.69\pm0.51$\\
\hline
\hline
\end{tabular}
 }
\end{table*}
%%%%%%%%%%%

%%%%%%%%
\begin{table*}[h!]\footnotesize
\caption{
Same as Table \ref{tab:fitotherScenI} except for yet-observed DCS  modes in Case I.
}
\label{tab:fitother2ScenI}
 \resizebox{\textwidth}{!} 
 {
\centering
\begin{tabular}
{ l | r rrr rr r
}
\hline
\hline
Channel&
$10^{3}\mathcal{B}$~~~~ &$\alpha$~~~~~~~&$\beta$~~~~~~~&$\gamma$~~~~~~
&$|A|$~~~ & $|B|$~~~ & $\delta_P-\delta_S$~~~\\
\hline
% DCS
\multirow{2}{*}
{$\Lambda_c^{+} \rightarrow n K^{+}$}&$0.12\pm0.07$&$-0.93\pm0.20$&$0.34\pm0.59$&$0.12\pm0.40$&$0.16\pm0.05$&$0.35\pm0.15$&$2.80\pm0.63$\\&$0.13\pm0.02$&$-0.95\pm0.04$&$0.31\pm0.11$&$0.07\pm0.15$&$0.16\pm0.01$&$0.37\pm0.05$&$2.83\pm0.12$\\
\multirow{2}{*}{$\Xi_c^0 \rightarrow \Sigma^{-} K^{+}$}&$0.82\pm0.14$&$-0.73\pm0.12$&$0.68\pm0.14$&$0.05\pm0.16$&$0.43\pm0.05$&$1.25\pm0.16$&$2.40\pm0.18$\\&$0.81\pm0.02$&$-0.72\pm0.03$&$0.69\pm0.03$&$0.05\pm0.04$&$0.43\pm0.01$&$1.25\pm0.03$&$2.38\pm0.04$\\
\multirow{2}{*}{$\Xi_c^0 \rightarrow p \pi^{-}$}&$0.19\pm0.03$&$-0.04\pm0.15$&$-0.90\pm0.11$&$-0.43\pm0.24$&$0.15\pm0.03$&$0.52\pm0.07$&$-1.61\pm0.17$\\&$0.19\pm0.01$&$-0.05\pm0.11$&$-0.90\pm0.04$&$-0.43\pm0.08$&$0.15\pm0.01$&$0.52\pm0.02$&$-1.63\pm0.12$\\
\multirow{2}{*}{$\Xi_c^0 \rightarrow n \pi^0 $}&$0.09\pm0.02$&$-0.04\pm0.15$&$-0.90\pm0.11$&$-0.43\pm0.24$&$0.10\pm0.02$&$0.36\pm0.05$&$-1.61\pm0.17$\\&$0.09\pm0.01$&$-0.05\pm0.11$&$-0.90\pm0.04$&$-0.43\pm0.08$&$0.10\pm0.01$&$0.37\pm0.02$&$-1.63\pm0.12$\\
\multirow{2}{*}{$\Xi_c^0 \rightarrow n \eta$}&$0.58\pm0.05$&$-0.47\pm0.16$&$0.53\pm0.12$&$-0.70\pm0.03$&$0.19\pm0.01$&$1.10\pm0.05$&$2.29\pm0.28$\\&$0.63\pm0.05$&$-0.42\pm0.06$&$0.55\pm0.07$&$-0.72\pm0.06$&$0.20\pm0.02$&$1.15\pm0.06$&$2.23\pm0.10$\\
\multirow{2}{*}{$\Xi_c^0 \rightarrow n \eta'$}&$0.32\pm0.06$&$-0.43\pm0.33$&$0.63\pm0.15$&$-0.65\pm0.32$&$0.18\pm0.07$&$1.07\pm0.18$&$2.17\pm0.31$\\&$0.38\pm0.08$&$-0.40\pm0.07$&$0.52\pm0.18$&$-0.75\pm0.15$&$0.17\pm0.04$&$1.20\pm0.16$&$2.23\pm0.15$\\
\multirow{2}{*}{$\Xi_c^{+} \rightarrow n \pi^{+}$}&$0.55\pm0.09$&$-0.04\pm0.15$&$-0.90\pm0.11$&$-0.43\pm0.24$&$0.15\pm0.03$&$0.52\pm0.07$&$-1.61\pm0.17$\\&$0.56\pm0.04$&$-0.05\pm0.11$&$-0.90\pm0.04$&$-0.43\pm0.08$&$0.15\pm0.01$&$0.52\pm0.02$&$-1.63\pm0.12$\\
\multirow{2}{*}{$\Xi_c^{+} \rightarrow \Sigma^0 K^{+}$}&$1.21\pm0.21$&$-0.73\pm0.12$&$0.68\pm0.14$&$0.05\pm0.16$&$0.30\pm0.03$&$0.89\pm0.12$&$2.40\pm0.18$\\&$1.21\pm0.04$&$-0.72\pm0.03$&$0.69\pm0.03$&$0.06\pm0.04$&$0.30\pm0.01$&$0.88\pm0.02$&$2.38\pm0.04$\\
\multirow{2}{*}{$\Xi_c^{+} \rightarrow p \pi^0$}&$0.28\pm0.05$&$-0.04\pm0.15$&$-0.90\pm0.11$&$-0.43\pm0.24$&$0.10\pm0.02$&$0.36\pm0.05$&$-1.61\pm0.17$\\&$0.28\pm0.02$&$-0.05\pm0.11$&$-0.90\pm0.04$&$-0.43\pm0.08$&$0.10\pm0.01$&$0.37\pm0.02$&$-1.63\pm0.12$\\
\multirow{2}{*}{$\Xi_c^{+} \rightarrow p \eta$}&$1.72\pm0.14$&$-0.47\pm0.16$&$0.53\pm0.12$&$-0.70\pm0.03$&$0.19\pm0.01$&$1.10\pm0.05$&$2.29\pm0.28$\\&$2.18\pm0.18$&$-0.49\pm0.05$&$0.54\pm0.12$&$-0.68\pm0.08$&$0.23\pm0.03$&$1.23\pm0.05$&$2.31\pm0.15$\\
\multirow{2}{*}{$\Xi_c^{+} \rightarrow p \eta'$}&$0.95\pm0.18$&$-0.43\pm0.33$&$0.63\pm0.15$&$-0.65\pm0.32$&$0.18\pm0.07$&$1.07\pm0.18$&$2.17\pm0.31$\\&$1.16\pm0.27$&$-0.40\pm0.08$&$0.50\pm0.18$&$-0.76\pm0.14$&$0.16\pm0.04$&$1.22\pm0.18$&$2.25\pm0.15$\\
\multirow{2}{*}{$\Xi_c^{+} \rightarrow \Lambda^0 K^{+}$}&$0.48\pm0.11$&$-0.17\pm0.18$&$-0.56\pm0.19$&$-0.81\pm0.15$&$0.08\pm0.03$&$0.70\pm0.09$&$-1.86\pm0.28$\\&$0.48\pm0.04$&$-0.20\pm0.10$&$-0.55\pm0.09$&$-0.81\pm0.06$&$0.08\pm0.02$&$0.70\pm0.03$&$-1.92\pm0.18$\\
\hline
\hline
\end{tabular}
 }
\end{table*}
%%%%%%%%%%%%%

%%%%%%%%%%%%%%%%
\begin{table*}[h!]\footnotesize
\caption{
Same as Table \ref{tab:fitother2ScenI} except for yet-observed DCS  modes in Case II.
}
\label{tab:fitother2ScenII}
 \resizebox{\textwidth}{!} 
 {
\centering
\begin{tabular}
{ l | r rrr rr r
}
\hline
\hline
Channel&
$10^{3}\mathcal{B}$~~~~ &$\alpha$~~~~~~~&$\beta$~~~~~~~&$\gamma$~~~~~~
&$|A|$~~~ & $|B|$~~~ & $\delta_P-\delta_S$~~~\\
\hline
%DCS
\multirow{2}{*}
{$\Lambda_c^{+} \rightarrow n K^{+}$}&$0.12\pm0.07$&$-0.88\pm0.25$&$0.42\pm0.54$&$0.23\pm0.38$&$0.16\pm0.04$&$0.32\pm0.14$&$2.70\pm0.61$\\&$0.12\pm0.01$&$-0.87\pm0.07$&$0.43\pm0.11$&$0.24\pm0.12$&$0.16\pm0.01$&$0.32\pm0.03$&$2.68\pm0.13$\\
\multirow{2}{*}{$\Xi_c^0 \rightarrow \Sigma^{-} K^{+}$}&$0.82\pm0.13$&$-0.75\pm0.10$&$0.66\pm0.12$&$0.03\pm0.13$&$0.43\pm0.04$&$1.26\pm0.13$&$2.42\pm0.16$\\&$0.82\pm0.02$&$-0.75\pm0.03$&$0.66\pm0.03$&$0.04\pm0.04$&$0.43\pm0.01$&$1.26\pm0.03$&$2.42\pm0.04$\\
\multirow{2}{*}{$\Xi_c^0 \rightarrow p \pi^{-}$}&$0.18\pm0.03$&$-0.07\pm0.24$&$-0.90\pm0.05$&$-0.44\pm0.09$&$0.14\pm0.01$&$0.51\pm0.05$&$-1.65\pm0.27$\\&$0.18\pm0.01$&$-0.07\pm0.10$&$-0.90\pm0.04$&$-0.44\pm0.07$&$0.14\pm0.01$&$0.51\pm0.02$&$-1.64\pm0.11$\\
\multirow{2}{*}{$\Xi_c^0 \rightarrow n \pi^0 $}&$0.09\pm0.02$&$-0.07\pm0.24$&$-0.90\pm0.05$&$-0.44\pm0.09$&$0.10\pm0.01$&$0.36\pm0.04$&$-1.65\pm0.27$\\&$0.09\pm0.01$&$-0.07\pm0.10$&$-0.90\pm0.04$&$-0.44\pm0.07$&$0.10\pm0.01$&$0.36\pm0.02$&$-1.64\pm0.11$\\
\multirow{2}{*}{$\Xi_c^0 \rightarrow n \eta$}&$0.54\pm0.03$&$-0.42\pm0.06$&$0.49\pm0.05$&$-0.76\pm0.03$&$0.17\pm0.01$&$1.08\pm0.03$&$2.28\pm0.11$\\&$0.52\pm0.04$&$-0.45\pm0.06$&$0.50\pm0.07$&$-0.74\pm0.05$&$0.17\pm0.02$&$1.06\pm0.05$&$2.30\pm0.10$\\
\multirow{2}{*}{$\Xi_c^0 \rightarrow n \eta'$}&$0.20\pm0.06$&$-0.42\pm0.14$&$0.27\pm0.16$&$-0.87\pm0.08$&$0.09\pm0.04$&$0.91\pm0.13$&$2.56\pm0.31$\\&$0.18\pm0.04$&$-0.44\pm0.07$&$0.36\pm0.24$&$-0.82\pm0.11$&$0.10\pm0.03$&$0.84\pm0.11$&$2.45\pm0.32$\\
\multirow{2}{*}{$\Xi_c^{+} \rightarrow n \pi^{+}$}&$0.55\pm0.10$&$-0.07\pm0.24$&$-0.90\pm0.05$&$-0.44\pm0.09$&$0.14\pm0.01$&$0.51\pm0.05$&$-1.65\pm0.27$\\&$0.55\pm0.04$&$-0.07\pm0.10$&$-0.90\pm0.04$&$-0.43\pm0.07$&$0.14\pm0.01$&$0.51\pm0.02$&$-1.64\pm0.11$\\
\multirow{2}{*}{$\Xi_c^{+} \rightarrow \Sigma^0 K^{+}$}&$1.21\pm0.19$&$-0.75\pm0.10$&$0.66\pm0.12$&$0.04\pm0.13$&$0.30\pm0.03$&$0.89\pm0.09$&$2.42\pm0.16$\\&$1.22\pm0.03$&$-0.75\pm0.03$&$0.66\pm0.03$&$0.04\pm0.04$&$0.30\pm0.01$&$0.89\pm0.02$&$2.42\pm0.04$\\
\multirow{2}{*}{$\Xi_c^{+} \rightarrow p \pi^0$}&$0.27\pm0.05$&$-0.07\pm0.24$&$-0.90\pm0.05$&$-0.44\pm0.09$&$0.10\pm0.01$&$0.36\pm0.04$&$-1.65\pm0.27$\\&$0.27\pm0.02$&$-0.07\pm0.10$&$-0.90\pm0.04$&$-0.44\pm0.07$&$0.10\pm0.01$&$0.36\pm0.02$&$-1.64\pm0.11$\\
\multirow{2}{*}{$\Xi_c^{+} \rightarrow p \eta$}&$1.60\pm0.09$&$-0.42\pm0.06$&$0.49\pm0.05$&$-0.76\pm0.03$&$0.17\pm0.01$&$1.08\pm0.03$&$2.28\pm0.11$\\&$1.90\pm0.11$&$-0.52\pm0.05$&$0.54\pm0.05$&$-0.66\pm0.05$&$0.22\pm0.01$&$1.14\pm0.04$&$2.33\pm0.08$\\
\multirow{2}{*}{$\Xi_c^{+} \rightarrow p \eta'$}&$0.61\pm0.18$&$-0.42\pm0.14$&$0.27\pm0.16$&$-0.87\pm0.08$&$0.09\pm0.04$&$0.91\pm0.13$&$2.56\pm0.31$\\&$0.52\pm0.13$&$-0.47\pm0.07$&$0.31\pm0.24$&$-0.83\pm0.10$&$0.09\pm0.02$&$0.84\pm0.12$&$2.55\pm0.36$\\
\multirow{2}{*}{$\Xi_c^{+} \rightarrow \Lambda^0 K^{+}$}&$0.46\pm0.08$&$-0.11\pm0.18$&$-0.59\pm0.12$&$-0.80\pm0.10$&$0.08\pm0.02$&$0.69\pm0.06$&$-1.76\pm0.29$\\&$0.46\pm0.04$&$-0.11\pm0.10$&$-0.58\pm0.09$&$-0.81\pm0.06$&$0.08\pm0.02$&$0.69\pm0.02$&$-1.75\pm0.17$\\
\hline
\hline
\end{tabular}
 }
\end{table*}
%%%%%%%%%%%

\vskip 5 pt
{\it 1.} We see from Table \ref{tab:resultScenI} that the predicted $\B(\Xi_c^0\to \Xi^0\eta')$  in Case I is more than three times bigger than the new Belle measurement, while $\B(\Xi_c^0\to \Xi^0\eta)$ is slightly larger than experiment. If we include the branching fractions  of $\Xi_c^0\to \Xi^0 \pi^0, \Xi^0\eta^{(')}$ measured by Belle as input, the amplitude $\tilde{E}_h$, which contributes to $\eta_1$ only, will be more constrained (see Tables \ref{tab:ampScenI} and \ref{tab:ampScenII}). Consequently,  the predicted $\B(\Xi_c^0\to \Xi^0\eta')$ is improved in Case II (c.f. Table  \ref{tab:resultScenII}). However, $\B(\Lambda_c^+\to\Sigma^+\eta')$ is reduced from $(0.39\pm 0.07)\%$ in Case I to  $(0.18\pm0.03)\%$ in Case II. The latter disagrees with the experimental value of $(0.41\pm0.08)\%$. Therefore, there is an issue of how to accommodate both $\Lambda_c^+\to\Sigma^+\eta'$ and $\Xi_c^0\to \Xi^0\eta'$ simultaneously. Further precise measurements of those modes with $\eta'$ are thus needed in the future.

\vskip 5 pt
{\it 2.}
The predictions based on the TDA and IRA generally agree with each other as it should be except for a few discrepancies mainly in the decay parameter $\alpha$,  for example, in $\Lambda_c^+\to p\pi^0$ in Case II, $\Xi_c^+\to \Sigma^+K_S$, $\Xi_c^0\to \Sigma^0 K_L$ and $\Xi_c^0\to\Lambda \pi^0$.

\vskip 5 pt
{\it 3.}
The recent LHCb measurements of the decay parameters $\alpha$ and $\beta$
for the decays $\Lambda_c^+\to\Lambda\pi^+$ and $\Lambda_c^+\to\Lambda K^+$ allow us to extract their phase shifts. From Table \ref{tab:LHCb} and Eq. (\ref{eq:phase}) we obtain
\begin{equation}
\label{eq:phaseshift}
(\delta_P - \delta_S)_{\Lambda_c^+\to\Lambda\pi^+} =2.70\pm 0.02~{\rm rad}, \qquad 
(\delta_P - \delta_S)_{\Lambda_c^+\to\Lambda K^+} =2.59\pm 0.17~{\rm rad}.
\end{equation}
Our results for the decay parameters $\alpha$ and $\beta$ are in agreement with experiment.
If the formula $\delta_P - \delta_S = \arctan({\beta}/{\alpha})$ is employed, it will yield $(\delta_P - \delta_S)_{\Lambda_c^+\to\Lambda\pi^+} =-0.44\pm 0.02$ rad and
$(\delta_P - \delta_S)_{\Lambda_c^+\to\Lambda K^+} =-0.55\pm 0.17$ rad. In this case, one needs to do manual adjustment to account for the $+\pi$ ambiguity. 

Likewise, the phase difference $\Delta$ between the two helicity amplitudes is related to the decay
parameters $\beta$ and $\gamma$ by
\begin{equation}
\Delta = 2 \arctan \frac{\beta}{\sqrt{\beta^2+\gamma^2}+\gamma}.
\label{eq:Delta}
\end{equation}
in analog to Eq. (\ref{eq:phase}).
The reader can check the validity of this relation using the LHCb measurements exhibited in Table \ref{tab:LHCb}.

\vskip 5 pt
{\it 4.}
For the decay $\Lambda_c^+\to \Xi^0 K^+$, BESIII found \cite{BESIII:XiK}
\begin{equation}
\alpha=0.01\pm0.16\pm0.03, \quad \beta=-0.64\pm0.69\pm0.13, \quad \gamma=-0.77\pm0.58\pm0.11,
\end{equation}
and uncovered two sets of solutions for the magnitudes of $S$- and $P$-wave amplitudes in units of $10^{-2} G_F\,{\rm GeV}^2$:
\begin{equation}
\label{eq:BESIIISol}
{\rm I}.~
\begin{cases}
 |A|=1.6^{+1.9}_{-1.6}\pm0.4\,,  \\  |B|=18.3\pm2.8\pm0.7\,,
 \end{cases} 
 \qquad
{\rm II}.~ \begin{cases}
|A|=4.3^{+0.7}_{-0.2}\pm0.4\,, \\ |B|=6.7^{+8.3}_{-6.7}\pm1.6\,,
\end{cases}
\end{equation}
as well as two solutions for the phase shift, 
\begin{equation}
\label{eq:phaseshifts}
\delta_P-\delta_S=-1.55\pm 0.25\pm0.05~~{\rm or}~~1.59\pm0.25\pm0.05~{\rm rad}.
\end{equation}
Our fits with $|A|=2.76\pm0.18$,  $|B|=9.71\pm0.47$, $\alpha_{\Xi^0K^+}=-0.04\pm0.12$, $\beta_{\Xi^0K^+}=-0.98\pm0.02$ and 
$\delta_P-\delta_S=-1.61\pm 0.12$ rad in Case I are consistent with the first phase-shift solution as well as the Lee-Yang parameters $\alpha_{\Xi^0K^+}$ and $\beta_{\Xi^0K^+}$. 
However, our fit of $\B(\Lambda_c^+\to \Xi^0 K^+)=(0.34\pm0.03)\%$ is smaller than the measured value of $(0.55\pm0.07)\%$ \cite{PDG}. Note that in our previous fits, $\delta_P-\delta_S=-2.15\pm0.65$ rad. Hence, our new fit of $-1.61\pm 0.12$ rad is improved substantially in precision and more close to the BESIII measurement.

It should be stressed that although the BESIII's measurement of $\alpha_{\Xi^0 K^+}$ is in good agreement with zero, it does not mean that the theoretical predictions in the 1990s with vanishing or very small $S$-wave amplitude are confirmed. 
We have checked that if we set $\delta_S^{X_i}=\delta_P^{X_i}=0$ from the outset and remove the input of $(\alpha_{\Xi^0K^+})_{\rm exp}$ , the fit $\alpha_{\Xi^0K^+}$ will be of order 0.95\,. Hence, we conclude that it is necessary to incorporate the phase shifts to accommodate the data. It is the smallness of $|\cos(\delta_P-\delta_S)|\sim 0.04$ that accounts for the nearly vanishing  $\alpha_{\Xi^0K^+}$.

\vskip 5pt
{\it 5.}
Besides the decay $\Lambda_c^+\to \Xi^0 K^+$, we have noticed  that  the following modes
$\Xi_c^0\to \Sigma^+ K^-, \Sigma^+\pi^-, p K^-, p\pi^-, n\pi^0$ and $\Xi_c^+\to p \pi^0, n\pi^+$ also receive contributions only from the topological $W$-exchange amplitude $\tilde{E}_1$. In the absence of strong phases in $S$- and $P$-wave amplitudes, they are expected to have large decay asymmetries. For example, $\alpha_{\Xi_c^0\to \Sigma^+K^-}$ was found to be $0.79^{+0.32}_{-0.33}$, $0.81\pm0.16$ and $0.98\pm0.20$, respectively, in Refs. \cite{Zhong:2022exp,Geng:2019xbo,Xing:2023dni}. Once the phase shifts are incorporated in the fit, the above-mentioned modes should have $\delta_P-\delta_S$ similar to that in $\Lambda_c^+\to \Xi^0 K^+$  and their decay asymmetries will become smaller (see Tables \ref{tab:resultScenI} and \ref{tab:resultScenII}).
In particular, for the CF channel $\Xi_c^0\to \Sigma^+ K^-$ whose branching fraction has been measured before, 
we predict that $\alpha_{\Xi_c^0\to \Sigma^+ K^-}=-0.04\pm0.12$ in Case I very similar to that of $\Lambda_c^+\to \Xi^0 K^+$. This can be used to test our theoretical framework.

\vskip 5pt
{\it 6.}
The predicted $\B(\Xi_c^0\to \Xi^-\pi^+)=(2.97\pm0.09)\%$ is noticeably higher than the measured value of $(1.80\pm0.52)\%$ by Belle \cite{Belle:2018kzz} and two times larger than the PDG value of $(1.42\pm0.27)\%$ \cite{PDG}. Nevertheless,  it is in a good agreement with the sum rule derived in both TDA and IRA, namely,
\begin{equation}
\label{eq:sumrule}
{\tau_{\Lambda_c^+}\over\tau_{\Xi_c^0}}\B(\Xi_c^0\to\Xi^-\pi^+) = 3\B(\Lambda_c^+\to \Lambda\pi^+)+\B(\Lambda_c^+\to \Sigma^0\pi^+)-{1\over \sin^2\theta_C}\B(\Lambda_c^+\to n\pi^+).
\end{equation}
This sum rule was first derived in Ref. \cite{Geng:2023pkr}. It is very useful to constrain the branching fraction of $\Xi_c^0\to\Xi^-\pi^+$. From the measured data collected in Table \ref{tab:expandave}, we find $\B(\Xi_c^0\to \Xi^-\pi^+)=(2.85\pm0.30)\%$ in good agreement with the aforementioned prediction.
This needs to be tested in the near future. 

\vskip 5pt
{\it 7.}
To guarantee the predictive reliability, all the fits in this work have been carried out  in both the TDA and IRA approaches, 
in addition to demonstrating  their equivalence.
Furthermore, we have considered two different cases, Case I and Case II, reflecting the current experimental status. Some inconsistencies occurred before,  such as  
the sign flip of $\beta(\Lambda_c^+ \to \Sigma^+ K_S)$  in  the TDA and IRA approaches have been corrected here. The consistency of the fitted results between the TDA and IRA approaches and the two cases reinforces our confidence in the reliability of most of the predictions. However, some slight discrepancies remain for certain observables in specific modes, such as $\alpha$ and $\gamma$ of the $\Lambda_c^+ \to p \pi^0$ mode in Case II, as shown in Table VI. These discrepancies highlight the need for additional experimental data to achieve more reliable predictions in the future.

\section{$CP$ violation}
In the charmed baryon sector, besides {\it CP} asymmetries in partial decay rates, two other 
{\it CP}-violation quantities of interest are:
\begin{equation}
\A_\alpha={ \alpha+\bar\alpha \over \alpha-\bar\alpha}, \qquad R_\beta={ \beta+\bar\beta \over \alpha-\bar\alpha}.
\end{equation}
They are related to weak and strong phase differences between $S$- and $P$-wave amplitudes, $\Delta\phi$ and $\Delta\delta\equiv \delta_P-\delta_S$, respectively, via
\begin{equation}
\A_\alpha=-\tan\Delta\delta\tan\Delta\phi, \qquad R_\beta=\tan\Delta\phi.
\end{equation}
{\it CP} asymmetries $\A_\alpha$ and $R_\beta$ for $\Lambda_c^+\to\Lambda \pi^+, \Lambda K^+$ and $\Lambda_c^+\to pK_S^0$ have been measured recently by LHCb with null results \cite{LHCb:2024tnq}.

The existence of strong phases in the partial-wave amplitudes of hadronic charmed baryon decays plays a pivotal role in a further exploration of {\it CP} violation in the charmed baryon decays.   
According to the standard model, {\it CP} violation is at a very small level in the decays of charmed hadrons. This is because of the relation of the CKM matrix elements, $\lambda_s\approx -\lambda_d$. As a consequence, {\it CP} violation in the charm sector is usually governed by $\lambda_b$ which is very tiny compared to $\lambda_d$ or $\lambda_s$ in magnitude. This also indicates that the corresponding QCD penguin and electroweak penguin are also rather suppressed.  In order to have direct {\it CP} asymmetries in partial rates, it is necessary to have nontrivial weak and strong phase differences.
We consider the singly-Cabibbo-suppressed charmed baryon decay with the amplitude $A=\lambda_d A_d+\lambda_s A_s$. Its {\it CP} asymmetry defined by
\begin{equation}
\A_{CP}\equiv { \Gamma(\B_c\to \B P)-\Gamma(\ov{\B}_c\to \bar{\B}\bar P)\over 
\Gamma(\B_c\to \B P)+\Gamma(\ov{\B}_c\to \bar{\B}\bar P) }
\end{equation}
has the expression
\begin{equation}
\begin{aligned}
\A_{CP} & =  {2 {\rm Im}(\lambda_d\lambda_s^*)\over |\lambda_d|^2} \,
{ {\rm Im}(A_dA_s^*)\over |A_d-A_s|^2}
 = 1.31\times 10^{-3} {|A_d A_s|\over |A_d-A_s|^2} \sin\delta_{ds}, \\
\end{aligned}
\end{equation}
where $\delta_{ds}$ is the strong phase of $A_s$ relative to $A_d$. If the decay amplitude is written as
$A=\frac12(\lambda_s-\lambda_d)(A_s-A_d)-\frac12\lambda_b(A_s+A_d)$, we will have
\begin{equation}
\begin{aligned}
\A_{CP} & = - {2 {\rm Im}[(\lambda_s-\lambda_d)\lambda_b^*]\over |\lambda_s-\lambda_d|^2} \,
{ {\rm Im}[(A_s-A_d)(A_s^*+A_d^*)\over |A_s-A_d|^2} \\
 & =  {4 {\rm Im}[(\lambda_s-\lambda_d)\lambda_b^*]\over |\lambda_s-\lambda_d|^2} { {\rm Im}(A_dA_s^*)\over |A_s-A_d|^2}= 1.31\times 10^{-3} {|A_d A_s|\over |A_s-A_d|^2} \sin\delta_{ds}, \\
\end{aligned}
\end{equation}
which is the same as before.
In general, {\it CP} asymmetry is expected to be of order $10^{-4}$ or even smaller. For recent studies in the approach based on $SU(3)$ flavor symmetry, see Refs. \cite{He:2024pxh,Xing:2024nvg,Sun:2024mmk}

In 2019 LHCb has announced the first observation of {\it CP} asymmetry difference between $D^0\to K^+K^-$ and $D^0\to \pi^+\pi^-$ at the per mille level, namely, $\Delta \A_{CP}\equiv \A_{CP}(K^+K^-)-\A_{CP}(\pi^+\pi^-)=(-1.54\pm0.29)\times 10^{-3}$ \cite{LHCb:CP}. In the standard-model estimate with the short-distance penguin contribution, we have the expression (see e.g. \cite{Cheng:2012a})
\begin{equation}
\Delta \A_{CP}\approx -1.31\times 10^{-3} \left( \left|{P\over T+E}\right|_{_{K\!K}}\sin\theta_{_{K\!K}}+ \left|{P\over T+E}\right|_{_{\pi\pi}}\sin\theta_{_{\pi\pi}} \right),
\end{equation}
where $\theta{_{K\!K}}$ is the strong phase of $(P/T)_{_{K\!K}}$ and likewise for $\theta{_{\pi\pi}}$. Since $|P/T|$ is na{\"i}vely expected to be of order $(\alpha_s(\mu_c)/\pi)\sim {\cal O}(0.1)$, it appears that $\Delta \A_{CP}$ is most likely of order $10^{-4}$ assuming strong phases close to $90^\circ$ or even less for realistic strong phases.  It was pointed out in Ref. \cite{Cheng:2012a} that there is a resonant-like final-state rescattering which has the same topology as the QCD-penguin. That is, the penguin topology receives sizable long-distance contributions through final-state interactions. In 2012 an ansatz that the long-distance penguin $P^{\rm LD}$ is of the same order of magnitude as $E$  was made in Ref. \cite{Cheng:2012a}. This ansatz of $P^{\rm LD}=E^{\rm LD}\approx E$ was justified later on by a  systematical
study of the final-state rescattering of the short-distance $T$ diagram in the topological diagrammatic approach \cite{Wang:2021rhd}.  Since the $W$-exchange topology can be extracted from the data, this implies that one can make a reliable prediction of $\Delta \A_{CP}$ which was carried out in Ref. \cite{Cheng:2012b} in 2012 with a prediction which is amazingly in excellent agreement with the later LHCb observation of 
{\it CP} violation in the charm meson sector in 2019. \footnote{Another prediction of $\Delta \A_{CP}$ based on pQCD \cite{Li:2012cfa}
also yielded  a similar numerical result.} 
Hence,  
it is the interference between tree and long-distance penguin amplitudes that pushes $\Delta \A_{CP}$ up to the $10^{-3}$ level \cite{Cheng:2012b}. 

By the same token, it is conceivable that direct {\it CP} asymmetry in the charmed baryon sector at the per mille level also can be achieved through final-state interactions as discussed recently in Ref. \cite{He:2024pxh}. Take the decay $\Lambda_c^+\to n\pi^+$ as an example. Its tilde TDA amplitude reads from Eqs. (\ref{eq:TDAtree}) and (\ref{eq:TDAlambdab2}) to be
\begin{equation}
A(\Lambda_c^+\to n\pi^+)  =\frac12(\lambda_s-\lambda_d)(-2\tilde T+\tilde C'+\tilde E_1)-\frac14\lambda_b(\tilde T+\tilde C+4\tilde b_4).
\end{equation}
Among the coefficients $\tilde b_1,\cdots,\tilde b_5$ in $\A^{\lambda_b}_{\rm TDA}$,  $\tilde b_5$ given by $\tilde T+\tilde C$ is related to the ${\bf 15^b}$ representation, while the rest  arises from the penguin-like or penguin contributions in the {\bf 3}. In principle, the short-distance contributions to $\tilde b_1,\cdots$ and $\tilde b_4$ can be calculated, but the LD parts are unknown.  Ignoring $\tilde b_4$ for the moment, it turns out that {\it CP} asymmetry in $\Lambda_c^+\to n\pi^+$ is very small. 
We find 
\begin{equation}
\A_{CP} =\begin{cases}   (-3.6\pm9.9)\times 10^{-6}, \cr ~~~(1.8\pm5.0)\times 10^{-6},
                 \end{cases} \qquad
\A_\alpha=\begin{cases}   (-5.4\pm4.7)\times 10^{-5}, & {\rm Case~I}, \cr
             (-1.3\pm1.1)\times 10^{-4}, & {\rm Case~II}.  \end{cases}
\end{equation}
Just as the case of charmed meson decays, final-state interactions of the tree topology can induce the penguin topology through the $s$-channel resonances and $t$- as well as $u$-channel rescattering. As discussed in Ref. \cite{He:2024pxh}, the LD contributions to the coefficients $\tilde f^b_3, \tilde f^c_3$ and $\tilde f^d_3$ in the IRA can be related to   $\tilde f^b, \tilde f^c, \tilde f^d$ and $\tilde f^e$ which are ready extracted from the global fits. Likewise, the coefficients $\tilde b_2, \tilde b_3$ and $\tilde b_4$ in the TDA can be related to    
$\tilde T, \tilde C, \tilde C'$ and $\tilde E_1$, in analog to the relation of $P^{\rm LD}=E$ in the $D$ meson case.  Of course, the situation becomes much more complicated in the charmed baryon sector.
In this manner, it is possible to have large {\it CP} violation at the level of $10^{-3}$. If we take the value of $\tilde b_4$, which is identical to $\tilde f^d_3$ in the IRA, from Table I of \cite{He:2024pxh}, we will obtain 
\begin{equation}
\A_{CP} =\begin{cases}   (-5.2\pm21.9)\times 10^{-4}, \cr (-4.4\pm22.4)\times 10^{-4},
                 \end{cases} \qquad
\A_\alpha=\begin{cases}   (-3.4\pm40.6)\times 10^{-4}, & {\rm Case~I}, \cr
             (-5.8\pm47.5)\times 10^{-4}, & {\rm Case~II}, \end{cases}
\end{equation}
where the large error in $\tilde b_4$ has been taken into account. Hence, both $\A_{CP}$ and $\A_\alpha$ can be enhanced to the $10^{-3}$ level through final-state rescattering.
In particular, large {\it CP} asymmetry differences
$\A_{CP}(\Xi_c^0\to pK^-)- \A_{CP}(\Xi_c^0\to \Sigma^+\pi^-)=(-3.68\pm0.37)\times 10^{-3}$ and $\A_\alpha(\Xi_c^0\to pK^-)- \A_\alpha(\Xi_c^0\to \Sigma^+\pi^-)=(1.17\pm0.37)\times 10^{-3}$ were predicted in Ref. \cite{He:2024pxh}.  This is very encouraging and our work along this line will be presented elsewhere. 

\section{Conclusions}

There exist two distinct ways in realizing the approximate SU(3) flavor symmetry of QCD to describe the two-body nonleptonic decays of charmed baryons: the irreducible SU(3) approach (IRA) and the topological diagram approach (TDA). The TDA has the advantage that it is more intuitive, graphic and easier to implement model calculations. The original expression of TDA amplitudes is given in Eq. (\ref{Eq:TDAamp}). Five independent TDA tree amplitudes are shown in Eq. (\ref{eq:tildeTDA}). In terms of the charmed baryon state $(\B_c)_i$ and the octet baryon $(\B_8)^i_j$, TDA amplitudes can be decomposed into $\A^{\rm tree}_{\rm TDA}$ (see Eq. (\ref{eq:TDAtree})) and $\A^{\lambda_b}_{\rm TDA}$ (Eqs. (\ref{eq:TDAlambdab}) and (\ref{eq:TDAlambdab2})), where the latter is proportional to $\lambda_b$. Likewise, $\mathcal{A}_{\rm IRAb}^{\rm tree}$ and $\mathcal{A}_{\rm IRAb}^{\lambda_b}$ in the IRA are displayed in Eqs. (\ref{eq:IRAb}) and (\ref{eq:IRAHe}), respectively. 

We perform  global fits to the currently available data of two-body charmed baryon decays based on the TDA and IRA at the tree level. The recent measurements of the decay parameters $\beta$ and $\gamma$ by LHCb enable to fix the sign ambiguity of $\beta$ and pick up the solution for $S$- and $P$-wave amplitudes.  
Phenomenological implications are as follows:

\begin{itemize}
\item
We consider two different global fits: Case I without the recent Belle data on $\Xi_c^0\to \Xi^0 \pi^0, \Xi^o\eta^{(')}$  and Case II with all the currently available data included. The difference mainly lies in the hairpin amplitude $\tilde E_h$ which is more constrained by the Belle data.
While the predicted $\B(\Xi_c^0\to \Xi^0\eta')$ is too large in Case I and improved in Case II, the fitted $\B(\Lambda_c^+\to\Sigma^+\eta')$ is too small in Case II and agrees with the data in Case I. Therefore, there is an issue of how to accommodate both $\Lambda_c^+\to\Sigma^+\eta'$ and $\Xi_c^0\to \Xi^0\eta'$ simultaneously, which needs to be clarified in the future. 

\item 
We urge to apply Eqs. (\ref{eq:phase}) and (\ref{eq:Delta}) to extract the phase shift $\delta_P-\delta_S$
from the decay parameters $\alpha$ and $\beta$
and the phase difference $\Delta$ between the two helicity amplitudes from $\beta$ and $\gamma$, respectively, to avoid a possible ambiguity of $\pm \pi$.

\item 
Since we use the new LHCb measurements to fix the sign ambiguity of $\beta$ and $\gamma$, the fit results for the phase shift $\delta_P-\delta_S$ and the magnitudes of $S$- and $P$-wave amplitudes are substantially improved over our previous analyses with smaller errors. The fitting 
branching fractions, decay parameters, magnitudes of $S$- and $P$-wave amplitudes and their phase shifts in both the TDA and IRA are presented in Tables \ref{tab:resultScenI}-\ref{tab:fitother2ScenII}. These results can be tested in the near future.

\item 
The number of the minimum set of tensor invariants in the IRA and the topological amplitudes in the TDA is the same, namely, five in the tree-induced amplitudes and four in the penguin amplitudes. 

\item
The predicted branching fraction of $(2.97\pm0.09)\%$ for $\Xi_c^0\to \Xi^-\pi^+$ is higher than its current value, but it is in a good agreement with the sum rule derived in both the TDA and IRA.  This should be tested soon.

\item 
To discuss direct \CP asymmetry in charmed baryon decays, we have shown that the amplitudes  $\A^{\lambda_b}_{\rm TDA}$ in the TDA  proportional to $\lambda_b$ are related to the penguin diagrams depicted in Fig. \ref{Fig:TopDiag} except one of them. In order to achieve \CP violation at the per mille level, we need to consider final-state rescattering of tree topologies which allows to generate the penguin topology. The work along this line will
be presented elsewhere.

\end{itemize}

%%%%%%%%%%%%%%%%%%%%%%%%%%%%%%%%%%%%%%%%%%%%%%%%%%
\section*{Acknowledgments}
%%%%%%%%%%%%%%%%%%%%%%%%%%%%%%%%%%%%%%%%%%%%%%%%%%

We are grateful to Chia-Wei Liu for valuable discussions. 
This research was supported in part by the National Science and Technology Council of R.O.C. under Grant No.~112-2112-M-001-026  and the National Natural Science Foundation of China under Grant Nos.~12475095 and U1932104.

\appendix
\section{Experimental Data}
All the currently available experimental data in hadronic decays of charmed baryons are collected in Table \ref{tab:expandave}. Although the absolute branching fractions of 
$\Xi_c^0\to \Xi^- K^+, \Lambda K_S^0, \Sigma^0 K_S^0$ and $\Sigma^+ K^-$ are available in the PDG \cite{PDG}, 
they are determined using $\B(\Xi_c^0\to \Xi^-\pi^+)=(1.43\pm0.27)\%$ as a benchmark, which is lower than the Belle's
measurement of $(1.80\pm0.52)\%$ \cite{Belle:2018kzz} and two times smaller than the theoretical expectation as discussed in this work. Therefore, we prefer to consider the ratios of the branching fractions of aforementioned four modes relative to the benchmark one $\Xi_c^0\to \Xi^-\pi^+$.

%----------Exprimental data----------
\begin{table}[tp!]
\footnotesize
\caption{Experimental data of branching fractions and decay asymmetries taken from PDG \cite{PDG}, BESIII \cite{BESIII:ppi0,BESIII:XiK,BESIII:2024cbr},  Belle \cite{Belle-II:2024jql} and LHCb \cite{LHCb:2024tnq}. For the decay $\Xi_c^0\to \Xi^-\pi^+$, we use the Belle's measurement of its branching fraction \cite{Belle:2018kzz} as input.
}
\vspace{-0.4cm}
\label{tab:expandave}
\begin{center}
\renewcommand\arraystretch{1}
%\resizebox{\textwidth}{!} 
{
\begin{tabular}
{ l c c c c c}
\hline \hline
Observable & PDG  & BESIII & Belle  & LHCb  & Average\\
\hline
$10^{2}\mathcal{B}(\Lambda_c^+\to\Lambda^0\pi^+)$&
$1.29\pm{0.05}$&&& & $1.29\pm{0.05}$\\

$10^{2}\mathcal{B}(\Lambda_c^+\to \Sigma^0 \pi^+)$&
$1.27\pm{0.06}$&&& & 
$1.27\pm{0.06}$\\

$10^{2}\mathcal{B}(\Lambda_c^+\to \Sigma^+ \pi^0)$&
$1.24\pm0.09$&&& &
$1.24\pm0.09$\\

$10^{2}\mathcal{B}(\Lambda_c^+\to \Sigma^+ \eta)$&
$0.32\pm0.05$&&& & 
$0.32\pm0.05$\\

$10^{2}\mathcal{B}(\Lambda_c^+\to \Sigma^+ \eta')$&
$0.41\pm0.08$&&& &
$0.41\pm0.08$\\

$10^{2}\mathcal{B}(\Lambda_c^+\to \Xi^0 K^+)$&
$0.55\pm0.07$&&&  & 
$0.55\pm0.07$\\

$10^{4}\mathcal{B}(\Lambda_c^+\to \Lambda^0 K^+)$&
$6.42\pm0.31$& & & &
$6.42\pm0.31$\\

$10^{4}\mathcal{B}(\Lambda_c^+\to \Sigma^0 K^+)$&
$3.70\pm0.31$& & & &
$3.70\pm0.31$\\

$10^{4}\mathcal{B}(\Lambda_c^+\to \Sigma^+ K_S)$&
$4.7\pm1.4$&& & &
$4.7\pm1.4$\\

$10^{4}\mathcal{B}(\Lambda_c^+\to n \pi^+)$&
$6.6\pm1.3$&&& &
$6.6\pm1.3$\\

$10^{4}\mathcal{B}(\Lambda_c^+\to p\pi^0)$&
$<0.8$& $1.79\pm0.41$ \cite{BESIII:2024cbr} & &  &
$1.79\pm0.41$\\

$10^{2}\mathcal{B}(\Lambda_c^+\to p K_S)$&
$1.59\pm0.07$&&& &
$1.59\pm0.07$\\

$10^{3}\mathcal{B}(\Lambda_c^+\to p \eta)$&
$1.57\pm0.12$ &  $1.63\pm0.33$ \cite{BESIII:ppi0} & &  & $1.58\pm0.11$ \\

$10^{4}\mathcal{B}(\Lambda_c^+\to p \eta')$&
$4.8\pm0.9$&&& &
$4.8\pm0.9$\\

$10^{2}\mathcal{B}(\Xi_c^0\to \Xi^- \pi^+)$&
$1.43\pm0.27$&& $1.80\pm0.52$ \cite{Belle:2018kzz}  & & 
$1.80\pm0.52$\\

$10^{2}\frac{\mathcal{B}(\Xi_c^0\to \Xi^- K^+)}{\mathcal{B}(\Xi_c^0\to \Xi^- \pi^+)}$&
$2.75\pm0.57$&&&& $2.75\pm0.57$\\
$10^{2}\frac{\mathcal{B}(\Xi_c^0\to \Lambda K_S^0)}{\mathcal{B}(\Xi_c^0\to \Xi^- \pi^+)}$&
$22.5\pm1.3$&&& & $22.5\pm1.3$\\
$10^{2}\frac{\mathcal{B}(\Xi_c^0\to \Sigma^0 K_S^0)}{\mathcal{B}(\Xi_c^0\to \Xi^- \pi^+)}$&
$3.8\pm0.7$&&& & $3.8\pm0.7$\\
$10^{2}\frac{\mathcal{B}(\Xi_c^0\to \Sigma^+ K^-)}{\mathcal{B}(\Xi_c^0\to \Xi^- \pi^+)}$&
$12.3\pm1.2$&&& & $12.3\pm1.2$\\
$10^{3}\mathcal{B}(\Xi_c^0\to\Xi^0\pi^0)$  &  & & $6.9\pm1.6$ \cite{Belle-II:2024jql}  & & $6.9\pm1.6$\\
$10^{3}\mathcal{B}(\Xi_c^0\to\Xi^0\eta)$ & & & $1.6\pm0.5$ \cite{Belle-II:2024jql} & & $1.6\pm0.5$\\
$10^{3}\mathcal{B}(\Xi_c^0\to\Xi^0\eta')$  & & & $1.2\pm0.4$ \cite{Belle-II:2024jql} & & $1.2\pm0.4$ \\
$10^{2}\mathcal{B}(\Xi_c^+\to \Xi^0 \pi^+)$&
$1.6\pm0.8$&&& & $1.6\pm0.8$ \\
$\alpha(\Lambda_c^+\to \Lambda^0 \pi^+)$& 
$-0.755\pm0.006$ & & & $-0.782\pm 0.010$ & $-0.762\pm0.006$  \\

$\alpha(\Lambda_c^+\to \Sigma^0 \pi^+)$&
$-0.466\pm0.018$ & & & & $-0.466\pm0.018$ \\

$\alpha(\Lambda_c^+\to p K_S)$&
$0.18\pm0.45$&&& $-0.744\pm0.015$  & $-0.743\pm0.028$ \\

$\alpha(\Lambda_c^+\to \Sigma^+ \pi^0)$&
$-0.484\pm0.027$ & & & & $-0.484\pm0.027$ \\

$\alpha(\Lambda_c^+\to \Sigma^+ \eta)$&
$-0.99\pm0.06$ & & & & $-0.99\pm0.06$  \\

$\alpha(\Lambda_c^+\to \Sigma^+ \eta')$&
$-0.46\pm0.07$ & & & & $-0.46\pm0.07$\\

$\alpha(\Lambda_c^+\to \Lambda^0 K^+)$&
$-0.585\pm0.052$ & & & $-0.569\pm0.065$ & $-0.579\pm0.041$ \\

$\alpha(\Lambda_c^+\to \Sigma^0 K^+)$&
$-0.54\pm0.20$ & & & & $-0.54\pm0.20$\\

$\alpha(\Lambda_c^+\to \Xi^0 K^+)$&
&$0.01\pm0.16$ \cite{BESIII:XiK} && & $0.01\pm0.16$
\\

$\alpha(\Xi_c^0\to \Xi^- \pi^+)$&
$-0.64\pm0.05$&&&& $-0.64\pm0.05$\\

$\alpha(\Xi_c^0\to\Xi^0\pi^0)$ & & & $-0.90\pm0.27$ \cite{Belle-II:2024jql}  & & $-0.90\pm0.27$\\

$\beta(\Lambda_c^+\to \Lambda^0 \pi^+)$& 
& & & $0.368\pm0.021$ & $0.368\pm0.021$ \\
$\beta(\Lambda_c^+\to \Lambda^0 K^+)$& 
& & & $0.35\pm0.13$ &  $0.35\pm0.13$\\

$\gamma(\Lambda_c^+\to \Lambda^0 \pi^+)$& 
& & & $0.502\pm0.017$ & $0.502\pm0.017$  \\
$\gamma(\Lambda_c^+\to \Lambda^0 K^+)$& 
& & & $-0.743\pm0.071$ & $-0.743\pm0.071$ \\
\hline \hline
\end{tabular}
}
\end{center}
\end{table}

%
% BibTeX or Biber users please use (the style is already called in the class, ensure that the "woc.bst" style is in your local directory)
% \bibliography{your_bib_file} % Replace "your_bib_file" with the actual name of your .bib file

\end{document}